\documentstyle[12pt]{article}
\def\lsim{\mathrel{\rlap {\raise.5ex\hbox{$ < $}}
{\lower.5ex\hbox{$\sim$}}}}
\def\gappeq{\mathrel{\rlap {\raise.5ex\hbox{$>$}}
{\lower.5ex\hbox{$\sim$}}}}
\def\lappeq{\mathrel{\rlap{\raise.5ex\hbox{$<$}}
{\lower.5ex\hbox{$\sim$}}}}

\begin{document}

\begin{titlepage}

\begin{flushright}
OUTP-97-12-P \\
hep-th/9703192 \\
\end{flushright}

\begin{centering}
\vspace{.05in}
{\Large {\bf Dilatonic Black Holes in Higher-Curvature
String Gravity II: Linear Stability  \\}}
 
\vspace{.4in}
{\bf P. Kanti}$^a$,  
{\bf N.E. Mavromatos}$^{b}$, {\bf J. Rizos}$^a$,\\  
{\bf K. Tamvakis}$^a$ and {\bf E. Winstanley}$^b$
\\
\vspace{.4in}
$^a$ Division of Theoretical Physics,
Physics Department,\\ University of Ioannina,
Ioannina GR 451 10, Greece, \\[3mm]
$^b$ University of Oxford, Department of Physics (Theoretical Physics),\\ 
1 Keble Road, Oxford. OX1 3NP U.K. \\

\vspace{.4in}
{\bf Abstract} \\
\vspace{.1in}
\end{centering}
{\small We demonstrate linear stability of 
the dilatonic Black Holes
appearing in a string-inspired higher-derivative  gravity theory 
with a Gauss-Bonnet curvature-squared term.
The proof is accomplished  by mapping the system to a one-dimensional
Schr\"odinger problem which admits no bound states. 
This result 
is important in that it constitutes a  
linearly stable example of a black hole that 
bypasses the `no-hair conjecture'. However, 
the dilaton hair is {\it secondary}
in the sense that is is not accompanied by any new quantum number
for the black hole solution.}

\vspace{1.0in}

\begin{flushleft} 
March 1997 \\
\end{flushleft} 

\end{titlepage}

\section{Introduction} 

In ref. \cite{kanti} we have presented analytic arguments
in favour of, and demonstrated numerically, 
the existence of dilatonic black hole solutions 
with non-trivial scalar hair, in a string-inspired 
higher-derivative gravity theory with a Gauss Bonnet 
(GB) curvature-squared term. The numerical solutions clearly 
demonstrated the existence of 
a regular event horizon and asymptotic flatness
of the four-dimensional spherically-symmetric space-time
configurations
considered in the analysis. There is a non-trivial dilaton 
(global) charge, which however is related to
the ADM mass of the black hole, and hence the hair 
is of secondary type~\cite{wilczek}: 
the gravitational field acts as a source for 
the scalar hair and one does not obtain a new 
independent set 
of quantum numbers characterizing the black hole\footnote{A 
similar situation occurs in the 
Higgs hair of the 
Einstein-Yang-Mills-Higgs systems~\cite{eymh,mw},
where the Yang-Mills field act as a source for the 
non-trivial configurations of the Higgs field
outside the horizon.}.  
Subsequent to the work of \cite{kanti}, other researchers 
confirmed these results by discussing the internal 
structure of the solutions behind the horizon,  
and demonstrating numerically 
the existence of curvature singularities~\cite{pomaz}.
Also an extension of the analysis of ref. \cite{kanti} 
to incorporate gauge fields 
became possible~\cite{maeda,kanti2}.

The key feature for the existence of such hairy black holes 
is the 
bypass of the no-scalar-hair 
theorems~\cite{beken} due to the fact that, as a result of the GB term, 
the scalar field stress tensor becomes {\it negative}
near the horizon~\cite{kanti}, thereby violating one of the 
main assumptions in the proof of the no-hair theorem~\cite{beken}. 
An additional element was the fact that the higher-derivative
GB term provides a sort of `repulsion' that balances the gravitational
attraction of the standard Einstein terms, and a black hole is formed.
In this respect
the GB term plays a similar r\^ole to the {\it non-Abelian}
gauge field kinetic terms 
in Einstein-Yang-Mills-Higgs theories~\cite{mw},
which are also notable exceptions of the no-scalar-(Higgs)-hair theorem. 

An important question which arises concerns the stability
of the dilatonic black holes. In the Yang-Mills case, the 
structures are similar to the sphaleron solutions of 
flat-space Yang-Mills theories, and thus unstable~\cite{mw}. 
This may be easily understood by the fact that the black hole
solutions owe their existence to a delicate balance between 
the gravitational attraction and the Yang-Mills repulsive forces.
On the other hand, the dilatonic black hole solutions are entirely
due to the existence of a {\it single} force, 
that of gravity. This already prompts one to think
that such structures might be stable.

It is the purpose of this article to argue that the 
dilatonic black holes are indeed stable under linear 
time-dependent perturbations of the classical solutions. 
To this end, we shall map the system of gravitational-dilaton
equations for spherically-symmetric solutions into a 
one-dimensional Schr\"odinger problem, where the
instabilities are equivalent to bound states. We shall prove
that our dilaton-graviton system 
admits {\it no bound states}. 
This result is important, since it constitutes an example of a hairy
black hole structure that appears to be, at least linearly,
stable. Its importance is 
also related to the fact that such higher-curvature gravity theories
are effective theories obtained from superstrings, which may imply that 
there is plenty of room in the gravitational sector of string theory 
to allow for physically sensible situations that are not covered 
by the no-hair theorem as stated~\cite{beken}. Unfortunately, at present  
non-linear stability of the dilaton-graviton-GB system cannot be 
checked analytically, and is left for future
investigations. 

\section{Relevant Formalism} 

We start by considering  
the action of the Einstein-Dilaton-Gauss-Bonnet (EDGB)
theory:
\begin{equation}
S=\int d^4 x \, \sqrt{-g} \left( \frac{R}{2} + \frac{1}{4}
\partial_{\mu} \phi \partial^{\mu} \phi + \frac{\alpha' e^\phi}{8 g^2}
{\cal R}^2_{GB} \right)
\label{1}
\end{equation}
where
\begin{equation}
{\cal R}^2_{GB}=R_{\mu\nu\rho\sigma}R^{\mu\nu\rho\sigma}-
4R_{\mu\nu}R^{\mu\nu}+R^2
\end{equation}
The spherically-symmetric ansatz for the metric takes the form
\begin{equation}
ds^2=e^{\Gamma(r,t)} dt^2-e^{\Lambda(r,t)} dr^2 -r^2(d\theta^2
+\sin ^2\theta \, d\varphi^2)
\label{2}
\end{equation}
The equations of motion derived from (\ref{1}) are:
\begin{eqnarray}
&~& \phi'' + \phi' 
\left( \frac{\Gamma'}{2} -\frac{\Lambda'}{2}+\frac{2}{r} \right)
-e^{\Lambda-\Gamma}
\left[ \ddot{\phi}+\frac{\dot{\phi}}{2}\,(\dot{\Lambda}
-\dot{\Gamma}) \right]~=~\nonumber \\
&~& 
 \frac{\alpha' e^\phi}{g^2 r^2} 
\left\{ \Gamma' \Lambda' e^{-\Lambda} 
- \dot{\Lambda}^2 e^{-\Gamma}
+\,\,(1-e^{-\Lambda})
\left[ \Gamma''+\frac{\Gamma'}{2}(\Gamma'-\Lambda')\right]
 \right. \nonumber \\
 &~& \left.
-(1-e^{-\Lambda})\,e^{\Lambda-\Gamma}\,
\left[ \ddot{\Lambda}
+\frac{\dot{\Lambda}}{2}\, (\dot{\Lambda}-\dot{\Gamma})\,
\right] \right\}
\label{3}
\end{eqnarray}  
\vspace{2mm} 
\begin{eqnarray}
\Lambda'
\left[ 1+ \frac{\alpha' e^\phi \phi'}{2g^2r}\, (1-3 e^{-\Lambda})
\right]
& = &  
\frac{r}{4}\left( \phi^{'2} + e^{\Lambda-\Gamma} \dot{\phi}^2
\right)
+\frac{1-e^\Lambda}{r}  \nonumber 
\\[3mm]
 &+& 
\frac{\alpha' e^\phi}
{g^2 r}\,(1-e^{-\Lambda}) 
\left[ \phi''  + \phi ^{'2} -
\frac{\dot{\phi} \dot{\Lambda}}{2}\, e^{\Lambda-\Gamma}
\right]
\label{4}
\end{eqnarray}
\vspace{2mm}
\begin{eqnarray}
\Gamma'
\left[ 1+ \frac{\alpha' e^\phi \phi'}{2g^2r}\,(1-3 e^{-\Lambda})
\right]
& = & 
\frac{r}{4}
\left( \phi^{'2} + e^{\Lambda-\Gamma} \dot{\phi}^2\right)
+ \frac{e^\Lambda-1}{r}  
\nonumber \\[3mm] 
&+&
\frac {\alpha' e^\phi}
{g^2 r}\,(1-e^{-\Lambda})\, e^{\Lambda-\Gamma} \, 
\left[ \ddot{\phi} +
\dot{\phi}^2 -\frac{\dot{\phi} \dot{\Gamma}}{2} \right]
\label{5}
\end{eqnarray}
\vspace{2mm}
\begin{equation}
\dot{\Lambda} 
\left[ 1+ \frac{\alpha' e^\phi \phi'}{2g^2r}\,(1-3 e^{-\Lambda})
\right]
=\frac{r \dot{\phi} \phi'}{2} + \frac{\alpha' e^\phi}{g^2 r}
\,(1-e^{-\Lambda}) 
\left( \dot{\phi} \phi'+ \dot{\phi}' 
-\frac{\dot{\phi} \Gamma'}{2} \right)
\label{6}
\end{equation}
%
\vspace{2mm}
\begin{eqnarray}
&~& 
\Gamma''
+\frac{\Gamma'}{2}\,(\Gamma'-\Lambda') 
+ \frac{(\Gamma'-\Lambda')}{r}
-e^{\Lambda-\Gamma} 
\left[ \ddot{\Lambda} + \frac{\dot{\Lambda}}{2}\,
(\dot{\Lambda}-\dot{\Gamma}) \right] ~=~
\nonumber \\
 & &
+\frac{1}{2} \,( e^{\Lambda-\Gamma}
\dot{\phi}^2-\phi^{'2})   
 +\frac{8\alpha'}{r}
 e^{-\Lambda}\,
\left\{ f' \Gamma'' + f'' \Gamma' + \frac{f' \Gamma'}{2}\,
(\Gamma'-3\Lambda') \right.
\nonumber \\ 
 & & \left.
+e^{\Lambda-\Gamma}\left[ \Lambda' \ddot{f}
-2 \dot{f}' \dot{\Lambda} +\frac{\dot{f}\dot{\Lambda}}
{2}\,(\Gamma'+\Lambda') 
-f'\ddot{\Lambda}
+(f'\dot{\Lambda}-\Lambda'\dot{f})
\,\frac{(\dot{\Gamma}+
\dot{\Lambda})}{2} \right] \right\}
\label{7}
\end{eqnarray} 
where $f=e^\phi/8 g^2$.\\[5mm]
For later use we note
that 
for the derivatives of the dilaton field at the horizon 
in the static case~\cite{kanti}
one has the following behaviour:  
\begin{equation} 
\phi_h'=\frac{g^2}{\alpha'}
r_he^{-\phi_h}\left(-1 \pm \sqrt{1-\frac{6(\alpha ')^2e^{2\phi_h}}{g^4r_h^4}}\right)
\label{phiprime}
\end{equation}
which implies that black hole solutions exists only if:
\begin{equation}
e^{\phi_h} \le \frac{g^2r_h^2}{\alpha'\sqrt{6}}
\label{condition}
\end{equation}
We now notice that~\cite{kanti} only one of the two branches of solutions
in (\ref{phiprime}), the one with the $+$ sign, leads to asymptotic
flatness of the fields,
and this is the branch we shall consider here. 
The equation for $\phi ''$ near $r_h$ is:
\begin{equation}
\phi ''=-\frac{1}{2}\frac{
(\frac{\alpha'}{g^2}e^\phi\phi'+ 2r)(6\frac{\alpha'}{g^2}e^\phi
+ \frac{\alpha'}{g^2}e^\phi \phi^{'2}r^2 + 2\phi'r^3)}
{-6\frac{\alpha^{'2}}{g^4}e^{2\phi} + 
\frac{\alpha'}{g^2}e^\phi \phi'r^3 + 2r^4}\,\Gamma ' + { O}(1) 
\label{phidoubleprime}
\end{equation}
which is finite ${ O}(1)$, as a result of (\ref{phiprime}).
The asymptotic form 
of the dilaton field and the
metric components near the even horizon $r \simeq r_h$ are:
\begin{eqnarray}
e^{-\Lambda(r)}&=& \lambda_1 (r-r_h)+\lambda_2 (r-r_h)^2 + ... \nonumber \\[4mm]
e^{\Gamma(r)}&=&\gamma_1 (r-r_h) +\gamma_2 (r-r_h)^2+... \nonumber \\[4mm]
\phi(r)&=&\phi_h+\phi'_h (r-r_h)+\phi''_h (r-r_h)^2+...
\label{asymptotia}
\end{eqnarray}
where: 
\begin{equation}
\lambda_1=2/(\alpha'e^{\phi_h}\phi_h'/g^2 + 2r_h)~,
\label{lambda1}
\end{equation}
and $\gamma_1$ 
is an arbitrary finite {\it positive} 
integration constant, which cannot be fixed by 
the equations of motion, since the latter involve only $\Gamma'(r)$
and not $\Gamma (r)$. This constant is fixed by the asymptotic
limit of the solutions at infinity.  
At infinity, one uses 
the following asymptotic behaviour: 
\begin{eqnarray}
e^{\Lambda(r)}&=&1+\frac{2M}{r}+\frac{16M^2-D^2}{4r^2}+
O \left( \frac{1}{r^3}\right)
 \nonumber \\[4mm]
e^{\Gamma(r)}&=&1-\frac{2M}{r}+O \left( \frac{1}{r^3} \right) 
\nonumber \\[4mm]
\phi(r)&=&\phi_\infty+\frac{D}{r}+\frac{MD}{r^2}+
O \left( \frac{1}{r^3} \right) 
\label{asinf2}
\end{eqnarray}
which guarantees asymptotic flatness of the space time. 
Above,  $M$ denotes the ADM mass of the black hole, and $D$ 
the dilaton charge. The numerical analysis 
of \cite{kanti} has shown that $M$ and $D$ are not independent
quantities, thereby leading to the secondary nature of the dilaton 
hair~\cite{kanti,wilczek}.

The black hole solutions of \cite{kanti} are characterized
uniquely by two parameters $(\phi_h, r_h)$. Note, however,
that the equations of motion remain invariant under
a shift $\phi~\rightarrow~\phi+\phi_0$ as long as it is accompanied
by a radial rescaling $r \rightarrow r e^{\phi_0/2}$. Due to the above
invariance it is sufficient to vary only one of $r_h$ and $\phi_h$.
In the present analysis we choose to keep $r_h$ fixed ($r_h=1)$,
and to vary $\phi_h$. A typical family of solutions has dilaton 
configurations (outside the horizon) of the form depicted in 
Figure 1. The solutions are characterized by negative $\phi_h$,
and {\it monotonic}, {\it non-intersecting} behaviour 
from $r_h$ until infinity. These are essential features
of the solutions, that we shall 
make use of in our linear stability analysis.

\section{Linear Stability Analysis}

We now consider perturbing the equations (\ref{3})-(\ref{7}) 
by time-dependent {\it linear} perturbations of the form: 
\begin{eqnarray}
 \Gamma (r,t) &=& \Gamma (r) + \delta \Gamma (r,t)=\Gamma (r) + \delta \Gamma (r)
e^{i\sigma t} \nonumber \\
 \Lambda (r,t) &=& \Lambda (r) + \delta \Lambda (r,t)=\Lambda (r) 
+ \delta \Lambda (r)
e^{i\sigma t} \nonumber \\
 \phi (r,t) &=& \phi (r) + \delta \phi (r,t)=\phi (r) + \delta \phi (r)
e^{i\sigma t} 
\label{perturbations}
\end{eqnarray}
where the variations $\delta\Gamma$, $\delta\Lambda$ and $\delta\phi$ are
assumed small (bounded), and the quantities without a $\delta$ prefactor denote classical time-independent solutions of the
equations (\ref{3})-(\ref{7}). The above {\it harmonic} time dependence
is sufficient for a {\it linear stability} analysis, since by assumption the 
linear variations are characterized 
by a well-defined
Fourier expansion in time $t$~\cite{mw,linear}. 
The linear stability analysis proceeds by mapping
the algebraic system of variations of the equations of motion under 
consideration to   
a stationary one-dimensional Schr\"odinger problem, in an appropriate 
potential well, 
in which 
the `squared frequencies' $\sigma ^2$ will constitute the energy eigenvalues. 
In the present problem, the `wavefunction' turns out
to be the dilaton linear variation
$\delta \phi (r) e^{i\sigma t}$.  
Instabilities, then, 
correspond to negative energy eigenstates (`bound states'),  
i.e. imaginary frequencies $\sigma$. As we shall show in this article,
for the system of variations corresponding to (\ref{3})-(\ref{7}),(\ref{perturbations}) the corresponding Schr\"odinger problem  admits {\it no bound states},
thereby proving the linear stability of the EDGB hairy black holes. 

As we shall discuss below, 
some technical complications arise, as usual~\cite{mw}, in the above process 
due to the fact that the `naive' stationary Schr\"odinger equation 
with respect to the original coordinate $r$ is not well defined
at some points of the domain of $r \in [r_h, \infty)$. This necessitates
a change of coordinates in such a way that the resulting Schr\"odinger problem 
is well defined. A convenient choice is provided by the so-called 
`tortoise' coordinate $r^*$~\cite{mw,linear}, which is defined in such a way
so that the domain $[r_h, \infty)$ is extended over the entire real axis
$r \rightarrow r^* \in (-\infty, \infty)$. In our specific problem,
we shall define the `tortoise' co-ordinate $r^{*}$ as~\cite{mw}:

\begin{equation}
\frac{dr^{*}}{dr}=e^{-(\Gamma-\Lambda)/2}
\label{tortoise}
\end{equation}
As we shall show, then, the associated stationary Schr\"odinger 
equation, pertaining to the dilaton variation in (\ref{perturbations}), 
will be 
of the form: 
\begin{equation}
 p_*^2u + {\cal V}(r^*)u(r^*) = -\sigma^2 u(r^*) \qquad ; \qquad p_* \equiv
\frac{d}{d r^*}
\label{tortoiseeq}
\end{equation}
where $u(r^*)$ is related to the dilaton variation $\delta \phi (r)$
in (\ref{perturbations}), and 
the potential ${\cal V}$ is well defined over the domain of validity
of $r^*$. 

Let us now proceed with our analysis.  
The perturbed equations (\ref{3})-(\ref{7}), under the variations
(\ref{perturbations}) read:
\vspace{4mm}
\begin{eqnarray}
&~&
\delta\phi''
+\delta\phi'\, \left( \frac{\Gamma'}{2} -\frac{\Lambda'}{2}+
\frac{2}{r}\right)
- \delta\phi\, \left[ \phi'' + \phi'\,
\left( \frac{\Gamma'}{2}-\frac{\Lambda'}{2} + \frac{2}{r}
\right) \,\right] -
e^{\Lambda-\Gamma} \delta \ddot{\phi} 
\nonumber \\[3mm] 
& & 
+\delta\Gamma' 
\left\{ \frac{\phi'}{2}- \frac{\alpha' e^{\phi}} 
{g^2 r^2}\, 
\left[  \Lambda' e^{-\Lambda} +(1-e^{-\Lambda})
\left( \Gamma'-
\frac{\Lambda'}{2}
\right) \right] \right\}
+\delta \Lambda\, \frac{\alpha'
e^{\phi}}{g^2 r^2}\, 
\left\{ 
\Gamma' \Lambda' e^{-\Lambda} \right.
\nonumber \\[3mm]
 & &
\left.-e^{-\Lambda} 
\left[\Gamma'' + \frac{\Gamma'}{2}\, (\Gamma'-\Lambda')
\right] \right\}- \delta\Lambda' \left\{ \frac{\phi'}{2} +
\frac{\alpha' e^{\phi}}{g^2 r^2} \left[ \Gamma' e^{-\Lambda} -
(1-e^{-\Lambda}) \frac{\Gamma'}{2}\right] \right\}
\nonumber \\[3mm]
 & & 
+\frac{\alpha'e^{\phi}}{g^2 r^2}\,(1-e^{-\Lambda})\, e^{\Lambda-\Gamma}
\delta \ddot{\Lambda}
-\frac{\alpha'e^{\phi}}{g^2
r^2}\,(1-e^{-\Lambda})\,\delta\Gamma''~=~0,
\label{8}
\end{eqnarray}
\vspace{5mm}
\begin{eqnarray}
&~& 
\delta\Lambda'\,
\left[ \,1+ \frac{\alpha' e^{\phi} \phi'}{2g^2r}\,(1-3
e^{-\Lambda})\,\right]
+\delta\phi'\,
\left[ \,-\frac{r \phi'}{2}+\frac{\alpha' e^{\phi}}{2g^2 r}\,
\Lambda' \,(1-3e^{-\Lambda})\,\right]
\nonumber \\[3mm]
 & &
+ \delta\Lambda 
\left\{ \,\frac{e^\Lambda}{r} +
\frac{\alpha' e^{\phi-\Lambda}}{g^2 r}\,\left[\frac{3 \phi'\Lambda'}{2} 
-(\phi''+\phi'^2)\right]\right\} 
+\delta\phi \frac{\alpha' e^{\phi}}{2g^2 r}\phi' \Lambda' (1-3
e^{-\Lambda})
\nonumber \\[3mm]
 & & 
-\frac{\alpha' e^{\phi}}{g^2 r}\,
(1-e^{-\Lambda})\,[\delta\phi''+2\phi'\,\delta\phi'+\delta\phi
(\phi''+\phi'^2)\,]~=~0,
\label{9}
\end{eqnarray}
\vspace{5mm}
\begin{eqnarray}
&~& 
\delta\Gamma'\,
\left[ \,1+ \frac{\alpha' e^{\phi} \phi'}{2g^2r}\,(1-3
e^{-\Lambda})\,\right]
+ \delta\Lambda \,\left[ -\frac{e^\Lambda}{r} + 
\frac{3\alpha' e^{\phi}}{2 g^2 r}\,\phi'\Gamma' e^{-\Lambda}\,
\right] 
\nonumber \\[3mm]
 & &
+\delta\phi \frac{\alpha' e^{\phi}}{2g^2 r}\phi' \Gamma' (1-3
e^{-\Lambda}) + 
\delta\phi'\,
\left[ \,-\frac{r \phi'}{2}+\frac{\alpha' e^{\phi}}{2g^2 r}\,
\Gamma' \,(1-3e^{-\Lambda})\,
\right] 
\nonumber \\[3mm]
& &
- \frac{\alpha' e^{\phi}}{g^2 r}\,
(1-e^{-\Lambda})\,e^{\Lambda-\Gamma}\,\delta\ddot{\phi}~=~0
\label{10}
\end{eqnarray}
\vspace{5mm}
\begin{equation}
\delta\dot{\Lambda} \,
\left[ 1+ \frac{\alpha' e^{\phi} \phi'}{2g^2r} \,(1-3
e^{-\Lambda}) \right] 
=\frac{r \phi'}{2}\,\delta\dot{\phi} +
\frac{\alpha' e^\phi}{g^2 r} \,(1-e^{-\Lambda}) \,
\left( \delta\dot{\phi} \,\phi'+ \delta\dot{\phi'} -
\delta\dot{\phi}\,\frac{\Gamma'}{2}\right)
\label{11}
\end{equation}
\vspace{5mm}
\begin{eqnarray}
&~&
\delta\Gamma''
\left( 1-\frac{\alpha' e^{\phi-\Lambda}}{g^2r}\,\phi'
\right) +
\frac{\delta\Gamma'}{2}\,(\Gamma'-\Lambda') 
+(\delta\Gamma'-
\delta\Lambda') \left( \frac{\Gamma'}{2}+\frac{1}{r}
\right)
-e^{\Lambda-\Gamma}\delta\ddot{\Lambda}
\nonumber \\[3mm]
  & & 
-\frac{\alpha' e^{\phi-\Lambda}}{g^2r}
\left\{\delta\phi'\,\Gamma''+ (\delta\phi''+2\phi'\,\delta\phi')\Gamma'+   
\delta\Gamma'(\phi''+\phi'^2)
 + \frac{\delta\phi'
\Gamma'}{2}\,(\Gamma'-3\Lambda') 
\right.
\nonumber \\[3mm]
 & &
 \left.
 + \frac{\phi' \delta\Gamma'}{2}\, 
(\Gamma'-3 \Lambda')
+\frac{\phi' \Gamma'}{2}\,(\delta\Gamma'-
3 \delta\Lambda') 
+ e^{\Lambda-\Gamma}\,(\Lambda' \delta\ddot{\phi}-
\phi' \delta\ddot{\Lambda})\,\right\} +\phi'\,\delta\phi'
\nonumber \\[3mm]
 & & 
-\frac{\alpha' e^{\phi-\Lambda}}{g^2r}
\,\left[ \,\phi'\Gamma'' + \Gamma'\,(\phi''+ \phi'^2) 
+\frac{\phi' \Gamma'}{2}\,
(\Gamma'-3 \Lambda')\right](\delta\phi-\delta\Lambda)=0 
\label{12}
\end{eqnarray} 
\vspace{4mm}
We can integrate eq.(\ref{11}) to obtain
\vspace{1mm}
\begin{eqnarray}
\delta\Lambda \,
\left[ 1+ \frac{\alpha' e^{\phi} \phi'}{2g^2r} \,(1-3
e^{-\Lambda})\right]
 & = &   
\frac{\alpha' e^\phi}{g^2 r} \,(1-e^{-\Lambda}) \,
\left( \delta\phi \,\phi'+ \delta\phi' -
\delta\phi\,\frac{\Gamma'}{2}\right)
\nonumber \\[3mm]
 &+&
\frac{r \phi'}{2}\,\delta\phi + \mu (r)
\label{13}
\end{eqnarray}
where $\mu(r)$ is an arbitrary function of $r$ which can be set equal
to zero if we require that $\delta\Lambda=0$ 
when $\delta\phi=0$. An 
independent check of this assertion can 
be obtained as follows: first 
we differentiate (\ref{13}) with respect to $r$ and then use
eq.(\ref{9}), as well as the time-independent equations of motion.
Thus, 
we obtain the following differential equation for $\mu(r)$
\begin{equation}
\mu'(r) + \mu(r)\,(\frac{\Gamma'}{2}-\frac{\Lambda'}{2}+
\frac{1}{r})=0
\end{equation}
This can be integrated to give $\mu(r)\sim e^{(\Lambda-\Gamma)/2}/r$.
When $r\rightarrow r_h$, $e^{(\Lambda-\Gamma)/2}\rightarrow \infty$ and
$\mu \rightarrow \infty$, which is incompatible with the assumption of a small 
$\delta\Lambda(r)$
required for the linear stability analysis.
Rejecting this solution, we are left
with the trivial one, $\mu=0$. 
For calculational convenience we also set $\alpha'/g^2=1$, from now on, 
which 
is achieved by an appropriate rescaling of the dilaton field. 

Rearranging the above equations, one obtains, after a tedious but 
straightforward procedure,  
an equation for
$\delta\phi$ which has the following structure:
\begin{equation}
A\,\delta\phi''+ 2B\,\delta\phi' + C\,\delta\phi +
\sigma^2\,E\,\delta\phi=0
\label{14}
\end{equation}
where $A$, $B$, $C$, and $E$ are rather complicated functions of
$\phi$, $\phi'$, $\phi''$, $\Lambda$, $\Lambda'$, $\Gamma$, $\Gamma'$
and $\Gamma''$. In the limit $r \rightarrow r_h$ these coefficients
take the form
\begin{eqnarray}
A & = & \frac{2\,\sqrt{1-\frac{6 e^{2\phi_h}}{r_h^4}}}
{1+\sqrt{1-\frac{6 e^{2\phi_h}}{r_h^4}}}
+ O\,(r-r_h)
\label{A}
\\ [3mm]
B & = & \frac{\sqrt{1-\frac{6 e^{2\phi_h}}{r_h^4}}}
{1+\sqrt{1-\frac{6 e^{2\phi_h}}{r_h^4}}}\,\frac{1}{(r-r_h)}+O\,(1)
\label{B}
\\ [3mm]
C & = & \frac{2\,e^{2\phi_h}}{r_h^4\,(1+\sqrt{1-
\frac{6 e^{2\phi_h}}{r_h^4}}\,)}
\,\frac{1}{(r-r_h)^2}     
+O\left( \frac{1}{r-r_h} \right)
\label{C}
\\[3mm]
E & = & \frac{r_h\,\sqrt{1-\frac{6 e^{2\phi_h}}{r_h^4}}}{\gamma_1}
\,\frac{1}{(r-r_h)^2}
+ O\left( \frac{1}{r-r_h} \right)
\label{E}
\end{eqnarray}
where we have used the asymptotic behaviour (\ref{asymptotia})
near the event horizon.
 
On the other hand, when $r \rightarrow \infty$ we obtain
\begin{eqnarray}
A&=& 1+ \frac{e^{\phi_\infty}DM}{r^4} + O(\frac{1}{r^5}) 
\label{infA}
\\[4mm]
B&=& \frac{1}{r}+\frac{M}{r^2}+O(\frac{1}{r^5}) \label{infB} \\[4mm]
C&=& \frac{D^2}{2r^4}+O(\frac{1}{r^5}) \label{infC} \\[4mm]
E&=& 1 + \frac{4M}{r}+\frac{4M^2}{r^2}+ \frac{e^{\phi_\infty}DM}{r^4}+
O\left( \frac{1}{r^5}\right)
\label{asinf1}
\end{eqnarray}
where we have used the asymptotic behaviour (\ref{asinf2}) near
infinity.
  
As we can see from (\ref{A})--(\ref{E}) the coefficients
of the Schr\"odinger equation (\ref{14}) are not finite at
the boundary $r=r_h$, where the variation $\delta \phi$ is bounded.
As mentioned previously, to
arrive at a well-defined Schr\"odinger problem,
one can use the `tortoise' coordinate (\ref{tortoise}).
Then,
the perturbed equation for the dilaton field takes the form
\begin{equation}
 A \,\frac{d^2\delta\phi}{dr^{*2}}
+\left[ \,2B\,e^{(\Gamma-\Lambda)/2}-
\frac{A}{2}\frac{d(\Gamma-\Lambda)}{dr^*}\,
\right] \,\frac{d\delta\phi}{dr^*}+
 e^{\Gamma-\Lambda}(C+\sigma^2E)\,\delta\phi=0
\label{dphi2}
\end{equation}
or
\begin{equation}
{\cal{A}}\,\frac{d^2\delta\phi}{dr^{*2}}+2{\cal{B}}
\frac{d\delta\phi}{dr^*}+({\cal{C}}+\sigma^2{\cal{E}})\,\delta\phi=0
\label{dphi3}
\end{equation}
where
\begin{equation}
{\cal{A}}=A \quad,\quad {\cal{B}}=B\,e^{(\Gamma-\Lambda)/2}-
\frac{A}{4}\frac{d(\Gamma-\Lambda)}{dr^*},
\label{caldef}
\end{equation}
\vspace{3mm}
\begin{equation}
{\cal{C}}=e^{\Gamma-\Lambda} C \quad,\quad
{\cal{E}}=e^{\Gamma-\Lambda}E
\label{caldef2}
\end{equation}
 Note that near the horizon
\begin{eqnarray}
e^{(\Gamma-\Lambda)/2}&=&\sqrt{\gamma_1 \lambda_1}\,(r-r_h) +
O(r-r_h)^2 \\[4mm]
\frac{d(\Gamma-\Lambda)}{dr^*}&=&2 \sqrt{\gamma_1 \lambda_1}+
O(r-r_h)
\label{tortoisehor}
\end{eqnarray}
where $\lambda_1$ is given by (\ref{lambda1}). 
As a result, all the coefficients in eq. (\ref{dphi2}) are now
well-behaved near the horizon $r_h$. In order to eliminate the
term
proportional to $\delta\phi'$, we first divide eq. (\ref{dphi2}) by
${\cal{A}}$ and then we use the function
\begin{equation}
F=\exp \left( \,\,\int _{-\infty}^{r^*}
\frac{{\cal{B}}}{{\cal{A}}}dr^{*'}\right)
\label{defF}
\end{equation}

Then, the equation for $\delta\phi$ takes the form
\begin{equation}
p_*^2u+
\left[ \,\frac{{\cal C}}{\cal{A}}+\sigma^2 \frac{{\cal E}}{\cal{A}}
-\frac{{\cal{B}}^2}{{\cal{A}}^2} -
p_*\left(\frac{{\cal{B}}}{{\cal{A}}}\right)
\right] \,u=0
\label{finalequation}
\end{equation}
where
\begin{equation}
 p_* \equiv \frac{d}{dr^*}
\label{pstar}
\end{equation}
and
we have set $u=F\,\delta\phi$.

It is straightforward to see that
\begin{equation}
\frac{{\cal B}}{{\cal{A}}}\rightarrow 0 \quad
{\rm for}  \quad  r \rightarrow r_h \quad {\rm and} 
\quad r \rightarrow \infty
\label{bcBA}
\end{equation}
independently of $r_h$, $\phi_h$.
In addition
\begin{equation}
\frac{{\cal B}}{{\cal{A}}}\frac{dr^*}{dr}={\rm finite} \quad
{\rm for}  \quad  r \rightarrow r_h  
\label{bchorizon}
\end{equation}
Moreover, as shown in Fig. 2, the function ${\cal B}/{\cal A}$ 
is well-behaved over the entire domain outside the 
horizon, implying the integrability of the function $F$. 
Also, the quantity ${\cal E}/{\cal A}$
is finite and always {\it positive} outside the horizon 
of the numerical black-hole solutions
of \cite{kanti} (see Fig. 3).
It is also immediately seen that
the eigenfunction $u_0$, corresponding to 
the eigenvalue $\sigma=0$, 
can be constructed out of the difference of any two
curves
in Fig. 1. {}From the {\it monotonic} and 
{\it non-intersecting} nature of the various members 
of the family of the numerical solutions
of Fig. 1, then, one can conclude 
that $u_0$ has {\it no nodes} in the 
domain $r^* \in (-\infty,\infty)$, and that 
$p_*^2u_0/u_0=e^{\Gamma - \Lambda}[\frac{1}{2}(\Gamma ' - \Lambda ')u_0'
+ u_0'']/u_0$ 
is {\it finite}. 
This, together  
with 
the finite and  smooth form 
of ${\cal B}/{\cal A}$ (Fig. 2), 
implies, on account of (\ref{finalequation}),  
the finiteness of the   
coefficient ${\cal C}/{\cal A}$ 
outside the horizon, without the need 
for an explicit numerical computation.
Thus, 
equation (\ref{finalequation})
assumes the form of an ordinary Schr\"odinger
with {\it regular} coefficients over the entire
domain of $r^*$.

{}From (\ref{defF}), (\ref{bcBA})
one obtains on the boundaries: 
\begin{equation} 
  p_*u_0|_{r^*=\pm \infty}=\frac{{\cal B}}{{\cal A}}u_0|_{r^*=\pm\infty}
+ Fp_*\delta\phi|_{r^*=\pm\infty} 
\label{bv}
\end{equation}
The $\delta \phi$
remains bounded at $r^*=\pm \infty$.
{}From the asymptotic behaviour
(\ref{asinf2}) it becomes clear that, at the $r=\infty$
boundary,
${\cal B}/{\cal A} \rightarrow B/A \sim 1/r$, independently of $\phi_h$.
Thus, $F(r^* \rightarrow \infty)=e^{\int_{-\infty}^{r}
\frac{B}{A}dr} \sim r$ for $r \rightarrow \infty$.
Hence,
the first term in (\ref{bv}) becomes $\delta \phi_{\infty} +
{\cal O}(\frac{1}{r})$, whilst the second term vanishes as
$1/r$, for $r\rightarrow \infty$ (see (\ref{asinf2})).
Hence, at the $r=\infty$ boundary
the boundary values of $p_*u_0$ are proportional to those
of $u_0$:
\begin{equation}
p_*u_0|_{r^*=\infty} = \frac{u_0}{r}|_{r^*=\infty}
\label{bvinf}
\end{equation}
On the other hand, from (\ref{defF}) and
(\ref{bchorizon}) it becomes clear that
at the boundary $r^*=-\infty$ (horizon),
$F(r^*=-\infty)=1$. Moreover, 
$p_*u_0|_{r^*=-\infty}$
is given by:
\begin{equation}
p_*u_0 |_{r^*=-\infty} = (r-r_h)
(\phi^{(2)'}- \phi^{(1)'})|_{r=r_h}
= \left((r-r_h)\frac{\partial \phi'_h}{\partial
\phi_h}u_0\right) \left| _{r=r_h} \right.
\label{variation}
\end{equation}
since, in our construction, each family of solutions is
uniquely
characterized~\cite{kanti} by the value $\phi_h$ for fixed
$r_h$
(we remind the
reader that here we chose to
keep $r_h=1$ (fixed) and vary $\phi_h$).
{}From (\ref{phiprime})
one easily observes that for linear
variations, $\delta \phi \equiv \phi^{(2)}-\phi^{(1)}$,
the difference
$\phi_h^{(2)'}-\phi_h^{(1)'}$ is {\it finite}. Hence,
\begin{equation}
p_*u_0|_{r^*=-\infty} \rightarrow 0
\label{bvzero}
\end{equation}
  
We now discuss the $\sigma^2 <0$ unstable modes $u_{\sigma}$. 
It can be easily seen from (\ref{infA})-(\ref{asinf1})
that the asymptotic form of  
equation (\ref{finalequation}) at the $r^*=\infty$ boundary
reads:
\begin{equation}
p_*^2u_{\sigma} =-\sigma^2 u_{\sigma} 
\label{unstable} 
\end{equation}
Hence, the bound-state solution $u_{\sigma}$ 
behaves as~\footnote{Here, and in the following, 
we insist on bounded,  
or- at most- linearly divergent $u_{\sigma}$, 
at the boundaries $r^*=\pm\infty$. 
This is due to the fact that, since $u=F\delta \phi$, 
and $F$ is independent of $\sigma$ and at most linearly divergent
at $r^*=\infty$, then it is 
only for such a behaviour that the variation $\delta\phi$ remains
bounded, as required by the linear stability analysis.
An exponentially divergent $u_{\sigma} \sim e^r$
at the boundaries, would imply $\delta \phi \sim e^r/F$, and,
hence,  
is not acceptable.}:
\begin{equation}
u_{\sigma}(r^*=\infty)=e^{-|\sigma|r^*}\rightarrow 0 
\label{unstbv}
\end{equation}
On the other hand, from (\ref{A})-(\ref{E})
it is obvious that, 
on the horizon, equation (\ref{finalequation}) assumes the form
(for the case $r_h=1$):
\begin{equation}
p_*^2u_{\sigma} + k^2u_{\sigma}=0 
\end{equation}
\begin{equation}
k^2 \equiv \frac{2\gamma_1e^{2\phi_h}}{(1+\sqrt{1-6e^{2\phi_h}})
\sqrt{1-6e^{2\phi_h}}}
+ \sigma^2=k_0^2+\sigma^2
\label{eqhorison}
\end{equation}
{}From (\ref{eqhorison}) 
one can see two possibilities near the horizon of the black hole: 
\begin{itemize}
\item{(i)} The `total energy' is such that $
0 > \sigma^2 >-\frac{2\gamma_1e^{2\phi_h}}{(1+\sqrt{1-6e^{2\phi_h}})
\sqrt{1-6e^{2\phi_h}}}$. Taking into 
account the asymptotic form at $r^*=\infty$ 
(\ref{unstable}) 
one also observes that 
in this case  the spectrum of the respective
Schr\"odinger equation is {\it continuous} and {\it non degenerate}.
The general solution of the perturbation $u_\sigma$ 
{\it near the horizon} is, therefore, 
oscillatory (unbound state):
\begin{equation}
   u_{\sigma}^{\pm} \sim e^{\pm~ikr^*}  
\qquad r^* \sim -\infty
\label{oscillatory}
\end{equation}
Such a continuum of states {\it cannot exist} in our case
by continuity.
Indeed, due to the non-degenerate nature of the 
eigenvalue problem, 
the limiting case $\sigma \rightarrow 0$ 
should yield 
the solution $u_0$. However, in the limit 
$\sigma^2 \rightarrow 0$, and in terms of $r$, one obtains
\begin{equation}
u_{0}^{\pm} \sim {\rm cos}\left(\frac{e^{\phi_h}}
{(1-6e^{2\phi_h})^{1/4}}\,{\rm ln}(r-r_h) + \varphi_0\right)
\label{slimit}
\end{equation}
where $\varphi _0$ is a constant phase shift, and we have taken $u_0$ 
to be the real part of eq. (\ref{oscillatory}). Then, 
\begin{equation}
p_*u_0 \sim -k_0\, 
{\rm sin}\left(\frac{e^{\phi_h}}{(1-6e^{2\phi_h})^{1/4}}\ln(r-r_h) 
+ \varphi_0\right)
\end{equation}
The above result is not in agreement with eq. (\ref{bvzero}), thereby
contradicting the 
non-degenerate nature of these solutions, which, in turn, 
implies the absense 
of such solutions in the problem (\ref{finalequation}). 

\item{(ii)} This leaves one with the second possibility of 
a {\it discrete spectrum} of {\it bound states}, which would occur  
for: 
\begin{equation}
\sigma^2<-\frac{2\gamma_1e^{2\phi_h}}{(1+\sqrt{1-6e^{2\phi_h}})
\sqrt{1-6e^{2\phi_h}}} < 0 
\label{condbound}
\end{equation}
As we shall show 
below this is also {\it not realized} due to the special form of $u_0$. 
\end{itemize}
To this end, one first observes that 
such bound states would {\it vanish } exponentially at the 
$r^*=-\infty$ boundary:
\begin{equation}
u_{\sigma}(r^*=-\infty) \sim e^{|k|\,r^*}
\rightarrow 0 
\label{unstbvh}
\end{equation}
Thus, on account of (\ref{bvinf},\ref{bvzero},\ref{unstbv},\ref{unstbvh})
the Wronskian of any two solutions $u_1,u_2$
of the equation
(\ref{finalequation}) with $\sigma^2 \le 0$
vanishes at the boundaries:
\begin{equation}
W=(u_1p_*u_2-u_2p_*u_1)|_{r^*=\pm\infty}=0
\label{wronskian}
\end{equation}

To count the unstable gravitational modes
of the original problem, one needs to count the
nodes of the wave function $u$ of
the one-dimensional Schr\"odinger problem
(\ref{finalequation}).
Fortunately, this can be done without
detailed knowledge of the solutions. As we shall
discuss below, all one needs to
observe is the monotonic and non-intersecting nature of the
dilaton curves in Fig. 1.
To this end, one first observes that
a standard `node rule' for the discrete spectrum
of the equation (\ref{finalequation})
applies, which is a direct consequence of
Fubini's theorem of ordinary differential equations~\cite{fubini}.
This theorem can be stated as follows:
consider
two differential equations: 
\begin{equation}
u'' + 2p_1u' + q_1u = 0
\label{eq1}
\end{equation}
\begin{equation}
u'' + 2p_2u' + q_2u = 0
\label{eq2}
\end{equation}
If,
\begin{equation}
p_2' + p_2^2 - q_2 \le p_1' + p_1^2 - q_1
\label{nodecondition}
\end{equation}
throughout the interval $[a,b]$, 
then, between {\it any two consecutive zeroes of a solution of}
(\ref{eq1}), in the interval $[a, b]$, {\it there is at least one zero of a
solution of} (\ref{eq2}).
 
 In our case, we can apply this theorem for
 two-different eigenfunctions, $u_1$, $u_2$,
 corresponding to
 eigenvalues  $\sigma_1^2$ and $\sigma_2^2$
 of (\ref{finalequation}). The interval $[a,b]$ is 
the entire domain of validity of 
the solutions of (\ref{finalequation}), $(-\infty,\infty)$,
including the boundaries at infinity. 
In this case,
 \begin{equation}
 p_i=0~, \qquad
 q_i=\frac{{\cal C}}{\cal{A}}+\sigma_i^2 \frac{{\cal E}}{\cal{A}}
 -\frac{{\cal{B}}^2}{{\cal{A}}^2}
 -p_*\left(\frac{{\cal B}}{{\cal A}}\right),~~~~~i=1,2
 \label{specificqp}
 \end{equation}
 Then,
 the
 (sufficient) condition for a `node rule' (\ref{nodecondition})
reads simply:
\begin{equation}
\frac{{\cal E}}{{\cal A}}\sigma^2_2 \ge
\frac{{\cal E}}{{\cal A}}\sigma^2_1
\label{nodesigma}
\end{equation}
 The positivity 
 of ${\cal E}/{\cal A}$ (Fig. 3) 
 implies that the condition (\ref{nodesigma}) becomes 
simply:
 \begin{equation}
 \sigma^2_2 \ge
 \sigma^2_1
 \label{nodesigmaalone}
 \end{equation}
This special version of the 
theorem is known as Sturm's theorem~\cite{messiah}.
As a corollary of Fubini/Sturm's  theorem  
one obtains the standard `node rule' for the 
number of zeroes of the eigenfunctions in the 
discrete spectrum of bound states, 
according to which if the 
eigenfunctions are ranked in order of 
increasing energy, then, 
the $n-th$ eigenfunction has $n-1$ nodes 
(excluding the boundary zeroes)~\cite{messiah} (`node rule').

 Consider, now, the case where $\sigma_2$
 corresponds to the zero eigenvalue of
 (\ref{finalequation}),
 $\sigma_0=0$. 
 As can be seen from the numerical solution of Fig. 1,
 the {\it monotonic} and {\it non-intersecting} character
 of the dilaton curves in the entire domain outside the horizon
 implies that
 the solution $u_2=u_0$, which, as we have mentioned earlier, 
 can be constructed 
 out of the difference of two such solutions, 
 has {\it no nodes} in the domain $r^* \in (-\infty, \infty)$.
 Since any solution $u_n$
 from the {\it discrete set} of negative eigenvalues
 $\sigma_n^2 < 0$ (unstable modes) has at least two nodes at the 
 boundaries, according to Fubini/Sturm's theorem, 
 $u_0$ should have at least one
 in the domain $r^* \in (-\infty, \infty)$. 
 This contradicts the fact that $u_0$ is nodeless. Thus, 
 the only consistent situation is the one {\it without} such
 {\it negative-energy}
 modes. This, in turn, implies {\it linear stability}
 for the dilaton-GB black holes of ref. \cite{kanti}.
The reader might worry about the divergent boundary 
conditions of $u_0$ at $r^*=\infty$, which makes it 
{\it not an ordinary} eigenfunction of a 
Schr\"odinger problem. In the appendix we argue that this 
is not an obstacle. In fact, we 
present an explicit 
proof of 
the absence of bound states in our case, following the 
same spirit used in the proof of Fubini/Sturm's theorem.  
The crucial element, which allows the 
standard proof to go through, is the 
special boundary condition for the Wronskian (\ref{wronskian}),
which is valid for the entire spectrum of eigenfunctions
of (\ref{finalequation}) with $\sigma^2 \le 0$, including 
the non-standard one $u_0$. 

  The above considerations can be extended straightforwardly
  to the case where $r_h \rightarrow 0$. All the coefficients
  of (\ref{finalequation}) are still well-defined in this case,
  which implies that the stability in principle does not change.
  However,
  according to the analysis of ref. \cite{kanti}, the case $r_h
  \rightarrow 0$
  corresponds to a singular curvature scalar
  \begin{equation}
  R \sim \frac{2}{r_h^2} \qquad r_h \rightarrow 0
  \label{curvsing}
  \end{equation}
  which implies the absence of `particle-like' solutions
  in the EDGB system~\cite{kanti,galtsov}.
  {}From the condition for the existence of black hole solutions
  (\ref{condition}), one observes that the only consistent
  value of $\phi_h$ for $r_h \rightarrow 0$ is~\cite{kanti}
  $\phi_h \rightarrow -\infty$.
  The limit is taken in such a way that $6\alpha'e^\phi_h/g^2r_h^2=1$.
  This implies that in such a
  case the GB term in the action (\ref{1}) becomes irrelevant,
  and one is left with the standard Einstein term, which admits
  only Schwarzschild black holes, known to be stable.
  This stability is confirmed by the above smooth limit of the
  coefficients in (\ref{finalequation}) as $r_h \rightarrow 0$.
   
\section{Remarks and Outlook}
 
 Above we have demonstrated linear stability
 of the dilatonic black hole solutions in
 the EDGB system, found in ref. \cite{kanti}.
 This result is important, since it
 constitutes an
  example of a stable, albeit secondary, hair
 that bypasses the no-hair conjecture~\cite{mw,beken}.
 Non-linear stability of the EDGB system, however,
 although expected, still remains an open issue.
  
Before closing we would like to compare our semi-analytic
results on linear stability
with some remarks in favour of stability
by virtue of a catastrophe theory approach
made in ref. \cite{maeda}.
As usual~\cite{mw,maeda2}, catastrophe theory can only indicate
{\it relative changes} of stability,
and hence cannot constitute
a `proof' of stability.
In ref. \cite{maeda} a numerical solution was found
for the $(+)$--branch of solutions (\ref{phiprime})
which indicated the existence of a `turning point' (TP)
in the $r_h$--$M$
(or equivalently $\phi_h$--$M$)
graph. The TP occurs at the `critical point' for the existence
of black hole solutions, which is the point where the
black hole acquires a minimal mass, below which no
solution is found.
In the numerical solution of ref. \cite{maeda}
a continuation beyond this critical point emerged,
which
ends at a point (`singular point') where a singularity
appears in the square of the Riemann tensor, as well as
in $\phi_h''$.
 
 The part of the solution from the critical point to
 the singular
 point
 was argued in ref. \cite{maeda} to be
 {\it relatively unstable}, as compared to the regular branch
 discussed here.
 Such a change in stability manifests itself as a cusp in
 an appropriate catastrophe theory diagram~\cite{mw,maeda2}.
 In ref. \cite{maeda}
 such a diagram has been chosen to be the
 diagram of the thermodynamic entropy~\cite{thooft}
 versus the
 mass of the black hole.
  
  In our numerical solutions~\cite{kanti}, used here,
  we found no evidence for such a continuation
  of the solution after the critical point, of minimum mass,
  which in our analysis
  is also the end point, where the
  two branches of (\ref{phiprime}) meet
  (see Figure 3).
  Our black hole solutions are uniquely specified
  by the pair $(r_h,\phi_h)$, which was essential
  in our linear stability analysis above.
  Moreover, for finite $r_h$, the quantity $\phi_h''$
  is finite~\cite{kanti}.
In this respect we 
are in agreement with the results 
of 
ref. \cite{pomaz}, where a branching of solutions 
was found 
only {\it inside} the event horizon. These authors 
have  also given a graph of the entropy 
versus the mass of the black hole solution 
of \cite{kanti} outside the horizon, 
and found, as expected, a smooth curve, with no cusps. 
  In this article
  we have proven {\it analytically}
  the stability under
  linear perturbations
of this (unique)
branch of the black hole solutions, which in the $r_h-M$ graphs
of Figs. 4,5
terminates at the
minimum-mass critical point.
 Of course, this result
is in agreement with the relative stability of this branch
{\it suggested}
by the catastrophe
theory analysis of ref. \cite{maeda}, but, as we said, the form of their
numerical solution
appears to be different, as it continues beyond the
critical point. Nevertheless, we consider 
this as an {\it indication} for 
the stability 
of the solutions beyond the linear approximation. 
This, however, still remains an open issue.
 
 As a final remark we would like to mention that
 the effects of gauge fields on the dilatonic black hole GB solutions
 have been considered in ref. \cite{maeda,kanti2}. {}From the point
 of view of stability, one expects
 that, in the case of `coloured' black holes,
 involving non-Abelian gauge fields,
 {\it instabilities} occur in {\it both} the gauge and
 gravitational sectors of the solutions.
 Instabilities in
 the gauge sector are of sphaleron type~\cite{mw}.
 Those in the
 gravitational sector can be studied in a similar way as for
 coloured black holes in
 Einstein-Yang-Mills theories~\cite{mw}. One
 can go beyond linear stability analysis in such systems,
 by invoking catastrophe
 theory~\cite{mw,maeda2},
 which is capable of giving the relative
 stability of various branches of solutions for the coloured
 EDGB black holes~\cite{maeda,kanti2}.
 However, analytic methods can still be combined
 with the catastrophe theory approach~\cite{mw} in order to
 {\it count} the {\it unstable
 modes} in both sectors, gauge and gravitational,
 by invoking appropriate maps of the system
 of perturbations into one-dimensional stationary
 Schr\"odinger problems~\cite{mw,linear}.
 We hope to return to a
 detailed analytic study of these issues
 in a future publication.

\section*{Acknowledgements}

Three of us (P.K., J.R., and K.T.) acknowledge 
traveling support from the EU-TMR
Network `Beyond the Standard Model'. 
P.K. and K.T. wish to thank the Greek Ministry of Research 
and Technology for financial support (PENED 95). P.K.  
also thanks the
Department of (Theoretical)
Physics of Oxford University for the hospitality, and 
partial financial support, during the final stages 
of this work. 
The work of N.E.M. is supported by a P.P.A.R.C. (UK)
Advanced Research Fellowship. 
E.W. thanks Oriel College, Oxford, for financial support.

\section*{Appendix: Absence of bound states}

In this Appendix we prove the absence of bound states in the
problem (\ref{finalequation}) by following a Wronskian treatment for the
entire set of
eigenmodes with $\sigma^2 \le 0$. 
This justifies the validity  
of 
Fubini/Sturm's theorem in our case.
We stress that the crucial point in the proof is the
special boundary conditions of the Wronskian
(\ref{wronskian}). These allow a standard Wronskian treatment
to go through for the $u_0$ solution of (\ref{finalequation}),
despite its (linear) divergence at the
$r^*=\infty$ boundary, which makes it not an ordinary
eigenfunction of a Schr\"odinger problem.

To this end, 
we consider first the two solutions of  (\ref{finalequation}), 
$u_0$, and $u_b$ - the ground state, at the bottom 
of the discrete spectrum. Both of these 
have {\it no nodes} 
in the interior domain of $r^*$, excluding the boundaries
(the node structure of $u_b$
follows from the `node rule').  
We, then, employ properties 
of the Wronskian of the solutions as follows: 
first
we multiply each equation with the
 other eigenfunction. Next, we subtract the resulting system
of equations,
 and then, integrate it over the entire domain of $r^* \in
 (-\infty,\infty)$. In this way one  
 obtains, in a standard
 fashion~\cite{messiah}:
 \begin{equation}
 \Delta W|_{r^*=\pm \infty} =(\sigma_b^2 - \sigma_0^2)
 \int_{-\infty}^{\infty}dr^{*}
 \frac{{\cal E}}{{\cal A}}u_bu_0
 \label{wronsk2}
 \end{equation}
 where the left-hand-side denotes the change in the Wronskian
 between the two boundaries.
 {}From (\ref{wronskian}) this {\it vanishes}.
 Moreover, as we have mentioned previously, ${\cal E}/{\cal A}$
 is {\it positive definite} for the entire domain of $r^* \in
 (-\infty,\infty)$
 (Figure 3).
 Since 
 $u_b$, $u_0$ have {\it no nodes}
 in the domain $(-\infty, \infty)$, excluding the boundaries, 
one obtains from
 (\ref{wronsk2}) that the {\it only consistent} case is
 the degenerate one $\sigma_b^2=\sigma_0^2$. But $\sigma_0^2=0$,
 whilst $\sigma_b^2 <0$ by assumption; this implies a contradiction,
excluding  $\sigma_b$ from the spectrum.

One repeats the construction, using $u_0$ and 
any of the higher eigenfunctions of the discrete spectrum, 
$u_n$, corresponding to $\sigma_n^2 <0$. 
The change in the Wronskian between $-\infty$ and 
the first encountered zero of $u_n$, at $r^*=z_0$, is then given by:
\begin{equation}
 \Delta W|_{r^*=-\infty}^{z_0} = -u_0p_*u_n|_{z_0}=
(\sigma_n^2 - \sigma_0^2)
 \int_{-\infty}^{z_0}dr^{*}
 \frac{{\cal E}}{{\cal A}}u_nu_0
\label{change}
\end{equation}
Without loss of generality, one may assume that $u_n > 0$
in the interval $(-\infty, z_0)$. Then $p_*u_n (z_0)<0$. 
It is immediate to see that there is a contradiction in (\ref{change}). 
The middle part 
has the sign of $u_0$, whilst the right-hand-side 
has the opposite sign of $u_0$.  
The case of a zero of $u_n$, at a point 
$z_0$, 
such that 
$p_*u(z_0)=0$ is dealt with similarly. In that case the contradiction
lies in the fact that the left-hand  side vanishes, whilst the 
right-hand side is a negative number (for $u_n >0$ in the interval).  
These results exclude the possibility of bound-state
eigenfunctions with zeroes
in $(-\infty, \infty)$.

The above analysis, therefore, implies the {\it absence of negative 
energy modes} (bound states) in the problem (\ref{finalequation}),
which, in turn, leads to {\it linear stability} for the 
Dilaton-Gauss-Bonnet black holes of ref. \cite{kanti}.

\newpage
\begin{figure}
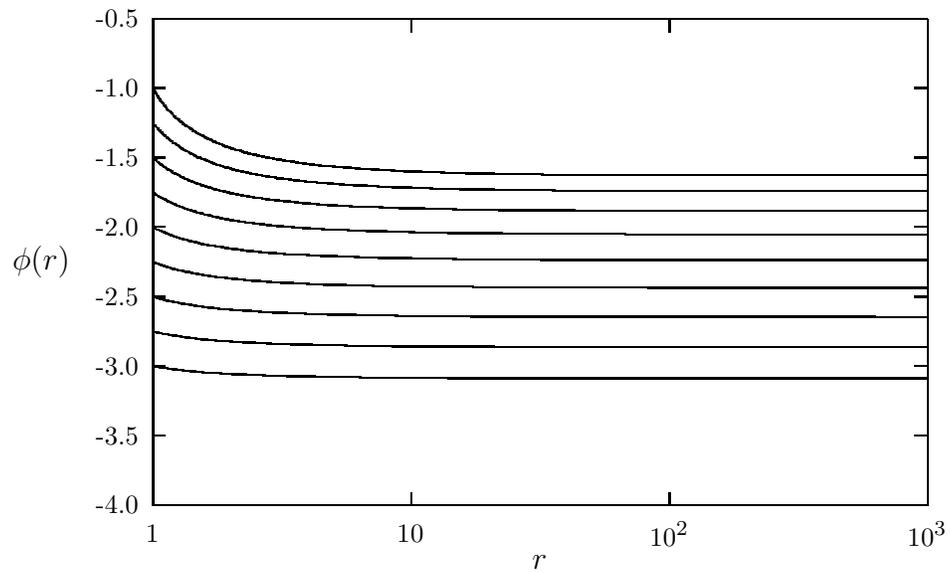

\begin{center}
\setlength{\unitlength}{0.240900pt}
\ifx\plotpoint\undefined\newsavebox{\plotpoint}\fi
\sbox{\plotpoint}{\rule[-0.200pt]{0.400pt}{0.400pt}}%

\caption{Dilaton configurations for  a family of black hole 
solutions, corresponding to fixed $r_h=1$, for various values of $\phi_h$.
Notice the monotonic behaviour of the solutions, and the 
fact that the curves do not intersect.}
\end{center}
\end{figure}

\begin{figure}
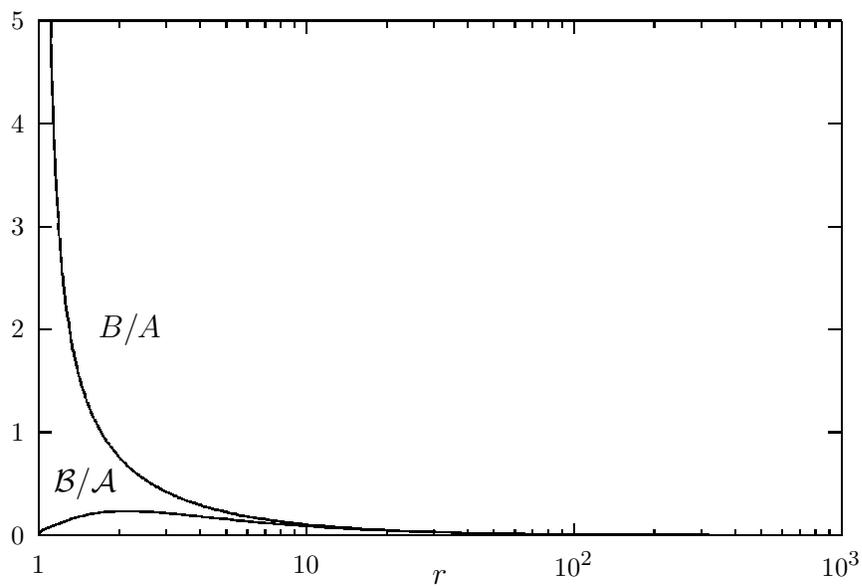

\begin{center}
\setlength{\unitlength}{0.240900pt}
\ifx\plotpoint\undefined\newsavebox{\plotpoint}\fi
\sbox{\plotpoint}{\rule[-0.200pt]{0.400pt}{0.400pt}}%
%
\caption{The graph depicts the coefficients $B/A$ and ${\cal B}/{\cal A}$,
for a typical member of the family of 
the black hole solutions of Fig.1 
corresponding to $\phi_h=-1$, $r_h=1$. 
It is clear that the coefficient ${\cal B}/{\cal A}$, incorporating 
the tortoise coordinate,  
is finite in the entire domain outside the horizon, thereby implying 
that the quantity $F$ is well-defined and integrable.}
\end{center}
\end{figure}
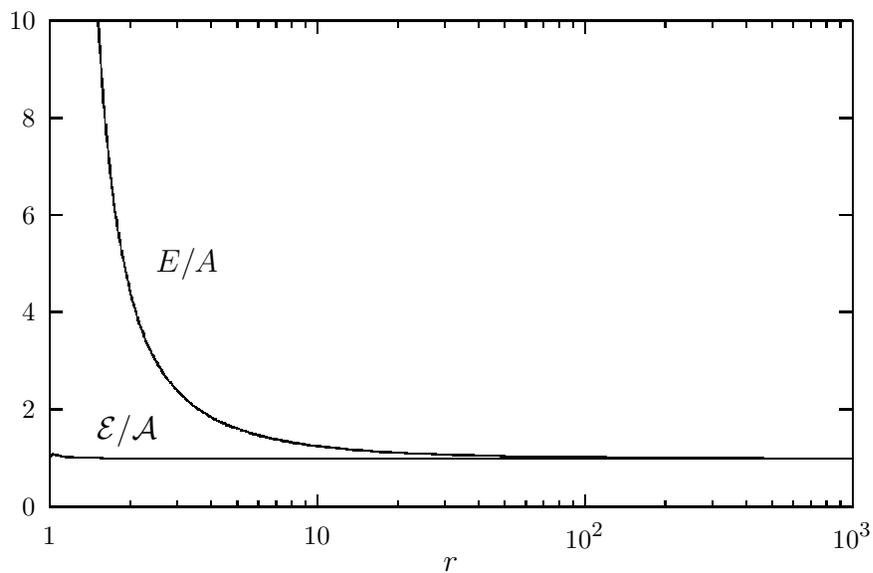
\begin{figure}
\begin{center}
\setlength{\unitlength}{0.240900pt}
\ifx\plotpoint\undefined\newsavebox{\plotpoint}\fi
\sbox{\plotpoint}{\rule[-0.200pt]{0.400pt}{0.400pt}}%
\begin{picture}(1500,900)(0,0)
\font\gnuplot=cmr10 at 10pt
\gnuplot
\sbox{\plotpoint}{\rule[-0.200pt]{0.400pt}{0.400pt}}%
\put(176.0,113.0){\rule[-0.200pt]{303.534pt}{0.400pt}}
\put(176.0,113.0){\rule[-0.200pt]{4.818pt}{0.400pt}}
\put(154,113){\makebox(0,0)[r]{0}}
\put(1416.0,113.0){\rule[-0.200pt]{4.818pt}{0.400pt}}
\put(176.0,266.0){\rule[-0.200pt]{4.818pt}{0.400pt}}
\put(154,266){\makebox(0,0)[r]{2}}
\put(1416.0,266.0){\rule[-0.200pt]{4.818pt}{0.400pt}}
\put(176.0,419.0){\rule[-0.200pt]{4.818pt}{0.400pt}}
\put(154,419){\makebox(0,0)[r]{4}}
\put(1416.0,419.0){\rule[-0.200pt]{4.818pt}{0.400pt}}
\put(176.0,571.0){\rule[-0.200pt]{4.818pt}{0.400pt}}
\put(154,571){\makebox(0,0)[r]{6}}
\put(1416.0,571.0){\rule[-0.200pt]{4.818pt}{0.400pt}}
\put(176.0,724.0){\rule[-0.200pt]{4.818pt}{0.400pt}}
\put(154,724){\makebox(0,0)[r]{8}}
\put(1416.0,724.0){\rule[-0.200pt]{4.818pt}{0.400pt}}
\put(176.0,877.0){\rule[-0.200pt]{4.818pt}{0.400pt}}
\put(154,877){\makebox(0,0)[r]{10}}
\put(1416.0,877.0){\rule[-0.200pt]{4.818pt}{0.400pt}}
\put(176.0,113.0){\rule[-0.200pt]{0.400pt}{4.818pt}}
\put(176,68){\makebox(0,0){1}}
\put(176.0,857.0){\rule[-0.200pt]{0.400pt}{4.818pt}}
\put(302.0,113.0){\rule[-0.200pt]{0.400pt}{2.409pt}}
\put(302.0,867.0){\rule[-0.200pt]{0.400pt}{2.409pt}}
\put(376.0,113.0){\rule[-0.200pt]{0.400pt}{2.409pt}}
\put(376.0,867.0){\rule[-0.200pt]{0.400pt}{2.409pt}}
\put(429.0,113.0){\rule[-0.200pt]{0.400pt}{2.409pt}}
\put(429.0,867.0){\rule[-0.200pt]{0.400pt}{2.409pt}}
\put(470.0,113.0){\rule[-0.200pt]{0.400pt}{2.409pt}}
\put(470.0,867.0){\rule[-0.200pt]{0.400pt}{2.409pt}}
\put(503.0,113.0){\rule[-0.200pt]{0.400pt}{2.409pt}}
\put(503.0,867.0){\rule[-0.200pt]{0.400pt}{2.409pt}}
\put(531.0,113.0){\rule[-0.200pt]{0.400pt}{2.409pt}}
\put(531.0,867.0){\rule[-0.200pt]{0.400pt}{2.409pt}}
\put(555.0,113.0){\rule[-0.200pt]{0.400pt}{2.409pt}}
\put(555.0,867.0){\rule[-0.200pt]{0.400pt}{2.409pt}}
\put(577.0,113.0){\rule[-0.200pt]{0.400pt}{2.409pt}}
\put(577.0,867.0){\rule[-0.200pt]{0.400pt}{2.409pt}}
\put(596.0,113.0){\rule[-0.200pt]{0.400pt}{4.818pt}}
\put(596,68){\makebox(0,0){10}}
\put(596.0,857.0){\rule[-0.200pt]{0.400pt}{4.818pt}}
\put(722.0,113.0){\rule[-0.200pt]{0.400pt}{2.409pt}}
\put(722.0,867.0){\rule[-0.200pt]{0.400pt}{2.409pt}}
\put(796.0,113.0){\rule[-0.200pt]{0.400pt}{2.409pt}}
\put(796.0,867.0){\rule[-0.200pt]{0.400pt}{2.409pt}}
\put(849.0,113.0){\rule[-0.200pt]{0.400pt}{2.409pt}}
\put(849.0,867.0){\rule[-0.200pt]{0.400pt}{2.409pt}}
\put(890.0,113.0){\rule[-0.200pt]{0.400pt}{2.409pt}}
\put(890.0,867.0){\rule[-0.200pt]{0.400pt}{2.409pt}}
\put(923.0,113.0){\rule[-0.200pt]{0.400pt}{2.409pt}}
\put(923.0,867.0){\rule[-0.200pt]{0.400pt}{2.409pt}}
\put(951.0,113.0){\rule[-0.200pt]{0.400pt}{2.409pt}}
\put(951.0,867.0){\rule[-0.200pt]{0.400pt}{2.409pt}}
\put(975.0,113.0){\rule[-0.200pt]{0.400pt}{2.409pt}}
\put(975.0,867.0){\rule[-0.200pt]{0.400pt}{2.409pt}}
\put(997.0,113.0){\rule[-0.200pt]{0.400pt}{2.409pt}}
\put(997.0,867.0){\rule[-0.200pt]{0.400pt}{2.409pt}}
\put(1016.0,113.0){\rule[-0.200pt]{0.400pt}{4.818pt}}
\put(1016,68){\makebox(0,0){10$^2$}}
\put(1016.0,857.0){\rule[-0.200pt]{0.400pt}{4.818pt}}
\put(1142.0,113.0){\rule[-0.200pt]{0.400pt}{2.409pt}}
\put(1142.0,867.0){\rule[-0.200pt]{0.400pt}{2.409pt}}
\put(1216.0,113.0){\rule[-0.200pt]{0.400pt}{2.409pt}}
\put(1216.0,867.0){\rule[-0.200pt]{0.400pt}{2.409pt}}
\put(1269.0,113.0){\rule[-0.200pt]{0.400pt}{2.409pt}}
\put(1269.0,867.0){\rule[-0.200pt]{0.400pt}{2.409pt}}
\put(1310.0,113.0){\rule[-0.200pt]{0.400pt}{2.409pt}}
\put(1310.0,867.0){\rule[-0.200pt]{0.400pt}{2.409pt}}
\put(1343.0,113.0){\rule[-0.200pt]{0.400pt}{2.409pt}}
\put(1343.0,867.0){\rule[-0.200pt]{0.400pt}{2.409pt}}
\put(1371.0,113.0){\rule[-0.200pt]{0.400pt}{2.409pt}}
\put(1371.0,867.0){\rule[-0.200pt]{0.400pt}{2.409pt}}
\put(1395.0,113.0){\rule[-0.200pt]{0.400pt}{2.409pt}}
\put(1395.0,867.0){\rule[-0.200pt]{0.400pt}{2.409pt}}
\put(1417.0,113.0){\rule[-0.200pt]{0.400pt}{2.409pt}}
\put(1417.0,867.0){\rule[-0.200pt]{0.400pt}{2.409pt}}
\put(1436.0,113.0){\rule[-0.200pt]{0.400pt}{4.818pt}}
\put(1436,68){\makebox(0,0){10$^3$}}
\put(1436.0,857.0){\rule[-0.200pt]{0.400pt}{4.818pt}}
\put(176.0,113.0){\rule[-0.200pt]{303.534pt}{0.400pt}}
\put(1436.0,113.0){\rule[-0.200pt]{0.400pt}{184.048pt}}
\put(176.0,877.0){\rule[-0.200pt]{303.534pt}{0.400pt}}
\put(806,23){\makebox(0,0){$r$}}
\put(343,495){\makebox(0,0)[l]{$E/A$}}
\put(250,232){\makebox(0,0)[l]{${\cal{E/A}}$}}
\put(176.0,113.0){\rule[-0.200pt]{0.400pt}{184.048pt}}
\put(252.17,854){\rule{0.400pt}{4.700pt}}
\multiput(251.17,867.24)(2.000,-13.245){2}{\rule{0.400pt}{2.350pt}}
\put(254.17,825){\rule{0.400pt}{5.900pt}}
\multiput(253.17,841.75)(2.000,-16.754){2}{\rule{0.400pt}{2.950pt}}
\multiput(256.61,795.25)(0.447,-11.625){3}{\rule{0.108pt}{7.167pt}}
\multiput(255.17,810.13)(3.000,-38.125){2}{\rule{0.400pt}{3.583pt}}
\put(259.17,748){\rule{0.400pt}{4.900pt}}
\multiput(258.17,761.83)(2.000,-13.830){2}{\rule{0.400pt}{2.450pt}}
\multiput(261.60,730.15)(0.468,-6.038){5}{\rule{0.113pt}{4.300pt}}
\multiput(260.17,739.08)(4.000,-33.075){2}{\rule{0.400pt}{2.150pt}}
\put(265.17,687){\rule{0.400pt}{3.900pt}}
\multiput(264.17,697.91)(2.000,-10.905){2}{\rule{0.400pt}{1.950pt}}
\multiput(267.61,667.21)(0.447,-7.607){3}{\rule{0.108pt}{4.767pt}}
\multiput(266.17,677.11)(3.000,-25.107){2}{\rule{0.400pt}{2.383pt}}
\put(270.17,636){\rule{0.400pt}{3.300pt}}
\multiput(269.17,645.15)(2.000,-9.151){2}{\rule{0.400pt}{1.650pt}}
\multiput(272.61,619.53)(0.447,-6.267){3}{\rule{0.108pt}{3.967pt}}
\multiput(271.17,627.77)(3.000,-20.767){2}{\rule{0.400pt}{1.983pt}}
\put(275.17,594){\rule{0.400pt}{2.700pt}}
\multiput(274.17,601.40)(2.000,-7.396){2}{\rule{0.400pt}{1.350pt}}
\multiput(277.61,580.30)(0.447,-5.151){3}{\rule{0.108pt}{3.300pt}}
\multiput(276.17,587.15)(3.000,-17.151){2}{\rule{0.400pt}{1.650pt}}
\put(280.17,559){\rule{0.400pt}{2.300pt}}
\multiput(279.17,565.23)(2.000,-6.226){2}{\rule{0.400pt}{1.150pt}}
\multiput(282.61,546.96)(0.447,-4.481){3}{\rule{0.108pt}{2.900pt}}
\multiput(281.17,552.98)(3.000,-14.981){2}{\rule{0.400pt}{1.450pt}}
\put(285.17,529){\rule{0.400pt}{1.900pt}}
\multiput(284.17,534.06)(2.000,-5.056){2}{\rule{0.400pt}{0.950pt}}
\multiput(287.61,518.62)(0.447,-3.811){3}{\rule{0.108pt}{2.500pt}}
\multiput(286.17,523.81)(3.000,-12.811){2}{\rule{0.400pt}{1.250pt}}
\put(290.17,503){\rule{0.400pt}{1.700pt}}
\multiput(289.17,507.47)(2.000,-4.472){2}{\rule{0.400pt}{0.850pt}}
\multiput(292.59,494.28)(0.482,-2.660){9}{\rule{0.116pt}{2.100pt}}
\multiput(291.17,498.64)(6.000,-25.641){2}{\rule{0.400pt}{1.050pt}}
\multiput(298.61,465.39)(0.447,-2.695){3}{\rule{0.108pt}{1.833pt}}
\multiput(297.17,469.19)(3.000,-9.195){2}{\rule{0.400pt}{0.917pt}}
\multiput(301.59,452.28)(0.477,-2.380){7}{\rule{0.115pt}{1.860pt}}
\multiput(300.17,456.14)(5.000,-18.139){2}{\rule{0.400pt}{0.930pt}}
\multiput(306.61,432.05)(0.447,-2.025){3}{\rule{0.108pt}{1.433pt}}
\multiput(305.17,435.03)(3.000,-7.025){2}{\rule{0.400pt}{0.717pt}}
\multiput(309.59,422.33)(0.482,-1.666){9}{\rule{0.116pt}{1.367pt}}
\multiput(308.17,425.16)(6.000,-16.163){2}{\rule{0.400pt}{0.683pt}}
\put(315.17,401){\rule{0.400pt}{1.700pt}}
\multiput(314.17,405.47)(2.000,-4.472){2}{\rule{0.400pt}{0.850pt}}
\multiput(317.59,396.71)(0.482,-1.214){9}{\rule{0.116pt}{1.033pt}}
\multiput(316.17,398.86)(6.000,-11.855){2}{\rule{0.400pt}{0.517pt}}
\put(323.17,380){\rule{0.400pt}{1.500pt}}
\multiput(322.17,383.89)(2.000,-3.887){2}{\rule{0.400pt}{0.750pt}}
\multiput(325.59,375.60)(0.477,-1.267){7}{\rule{0.115pt}{1.060pt}}
\multiput(324.17,377.80)(5.000,-9.800){2}{\rule{0.400pt}{0.530pt}}
\multiput(330.61,364.26)(0.447,-1.132){3}{\rule{0.108pt}{0.900pt}}
\multiput(329.17,366.13)(3.000,-4.132){2}{\rule{0.400pt}{0.450pt}}
\multiput(333.59,358.26)(0.477,-1.044){7}{\rule{0.115pt}{0.900pt}}
\multiput(332.17,360.13)(5.000,-8.132){2}{\rule{0.400pt}{0.450pt}}
\put(338.17,348){\rule{0.400pt}{0.900pt}}
\multiput(337.17,350.13)(2.000,-2.132){2}{\rule{0.400pt}{0.450pt}}
\multiput(340.59,344.60)(0.477,-0.933){7}{\rule{0.115pt}{0.820pt}}
\multiput(339.17,346.30)(5.000,-7.298){2}{\rule{0.400pt}{0.410pt}}
\put(345.17,335){\rule{0.400pt}{0.900pt}}
\multiput(344.17,337.13)(2.000,-2.132){2}{\rule{0.400pt}{0.450pt}}
\multiput(347.59,332.26)(0.477,-0.710){7}{\rule{0.115pt}{0.660pt}}
\multiput(346.17,333.63)(5.000,-5.630){2}{\rule{0.400pt}{0.330pt}}
\put(352.17,324){\rule{0.400pt}{0.900pt}}
\multiput(351.17,326.13)(2.000,-2.132){2}{\rule{0.400pt}{0.450pt}}
\multiput(354.60,321.09)(0.468,-0.774){5}{\rule{0.113pt}{0.700pt}}
\multiput(353.17,322.55)(4.000,-4.547){2}{\rule{0.400pt}{0.350pt}}
\put(358.17,315){\rule{0.400pt}{0.700pt}}
\multiput(357.17,316.55)(2.000,-1.547){2}{\rule{0.400pt}{0.350pt}}
\multiput(360.59,312.59)(0.477,-0.599){7}{\rule{0.115pt}{0.580pt}}
\multiput(359.17,313.80)(5.000,-4.796){2}{\rule{0.400pt}{0.290pt}}
\put(365,307.17){\rule{0.482pt}{0.400pt}}
\multiput(365.00,308.17)(1.000,-2.000){2}{\rule{0.241pt}{0.400pt}}
\multiput(367.59,304.51)(0.488,-0.626){13}{\rule{0.117pt}{0.600pt}}
\multiput(366.17,305.75)(8.000,-8.755){2}{\rule{0.400pt}{0.300pt}}
\multiput(375.00,295.94)(0.481,-0.468){5}{\rule{0.500pt}{0.113pt}}
\multiput(375.00,296.17)(2.962,-4.000){2}{\rule{0.250pt}{0.400pt}}
\multiput(379.00,291.93)(0.492,-0.485){11}{\rule{0.500pt}{0.117pt}}
\multiput(379.00,292.17)(5.962,-7.000){2}{\rule{0.250pt}{0.400pt}}
\multiput(386.00,284.94)(0.481,-0.468){5}{\rule{0.500pt}{0.113pt}}
\multiput(386.00,285.17)(2.962,-4.000){2}{\rule{0.250pt}{0.400pt}}
\multiput(390.00,280.93)(0.581,-0.482){9}{\rule{0.567pt}{0.116pt}}
\multiput(390.00,281.17)(5.824,-6.000){2}{\rule{0.283pt}{0.400pt}}
\multiput(397.00,274.95)(0.462,-0.447){3}{\rule{0.500pt}{0.108pt}}
\multiput(397.00,275.17)(1.962,-3.000){2}{\rule{0.250pt}{0.400pt}}
\multiput(400.00,271.93)(0.710,-0.477){7}{\rule{0.660pt}{0.115pt}}
\multiput(400.00,272.17)(5.630,-5.000){2}{\rule{0.330pt}{0.400pt}}
\put(407,266.17){\rule{0.700pt}{0.400pt}}
\multiput(407.00,267.17)(1.547,-2.000){2}{\rule{0.350pt}{0.400pt}}
\multiput(410.00,264.94)(0.920,-0.468){5}{\rule{0.800pt}{0.113pt}}
\multiput(410.00,265.17)(5.340,-4.000){2}{\rule{0.400pt}{0.400pt}}
\put(417,260.17){\rule{0.700pt}{0.400pt}}
\multiput(417.00,261.17)(1.547,-2.000){2}{\rule{0.350pt}{0.400pt}}
\multiput(420.00,258.94)(0.774,-0.468){5}{\rule{0.700pt}{0.113pt}}
\multiput(420.00,259.17)(4.547,-4.000){2}{\rule{0.350pt}{0.400pt}}
\put(426,254.17){\rule{0.700pt}{0.400pt}}
\multiput(426.00,255.17)(1.547,-2.000){2}{\rule{0.350pt}{0.400pt}}
\multiput(429.00,252.95)(0.909,-0.447){3}{\rule{0.767pt}{0.108pt}}
\multiput(429.00,253.17)(3.409,-3.000){2}{\rule{0.383pt}{0.400pt}}
\put(434,249.67){\rule{0.723pt}{0.400pt}}
\multiput(434.00,250.17)(1.500,-1.000){2}{\rule{0.361pt}{0.400pt}}
\multiput(437.00,248.95)(1.132,-0.447){3}{\rule{0.900pt}{0.108pt}}
\multiput(437.00,249.17)(4.132,-3.000){2}{\rule{0.450pt}{0.400pt}}
\put(443,245.67){\rule{0.482pt}{0.400pt}}
\multiput(443.00,246.17)(1.000,-1.000){2}{\rule{0.241pt}{0.400pt}}
\put(445,244.17){\rule{1.300pt}{0.400pt}}
\multiput(445.00,245.17)(3.302,-2.000){2}{\rule{0.650pt}{0.400pt}}
\put(451,242.17){\rule{0.482pt}{0.400pt}}
\multiput(451.00,243.17)(1.000,-2.000){2}{\rule{0.241pt}{0.400pt}}
\put(453,240.17){\rule{1.100pt}{0.400pt}}
\multiput(453.00,241.17)(2.717,-2.000){2}{\rule{0.550pt}{0.400pt}}
\put(458,238.67){\rule{0.723pt}{0.400pt}}
\multiput(458.00,239.17)(1.500,-1.000){2}{\rule{0.361pt}{0.400pt}}
\put(461,237.17){\rule{1.100pt}{0.400pt}}
\multiput(461.00,238.17)(2.717,-2.000){2}{\rule{0.550pt}{0.400pt}}
\multiput(468.00,235.94)(1.212,-0.468){5}{\rule{1.000pt}{0.113pt}}
\multiput(468.00,236.17)(6.924,-4.000){2}{\rule{0.500pt}{0.400pt}}
\put(477,231.67){\rule{1.204pt}{0.400pt}}
\multiput(477.00,232.17)(2.500,-1.000){2}{\rule{0.602pt}{0.400pt}}
\multiput(482.00,230.95)(1.579,-0.447){3}{\rule{1.167pt}{0.108pt}}
\multiput(482.00,231.17)(5.579,-3.000){2}{\rule{0.583pt}{0.400pt}}
\put(490,227.67){\rule{0.964pt}{0.400pt}}
\multiput(490.00,228.17)(2.000,-1.000){2}{\rule{0.482pt}{0.400pt}}
\put(494,226.17){\rule{1.700pt}{0.400pt}}
\multiput(494.00,227.17)(4.472,-2.000){2}{\rule{0.850pt}{0.400pt}}
\put(502,224.67){\rule{0.964pt}{0.400pt}}
\multiput(502.00,225.17)(2.000,-1.000){2}{\rule{0.482pt}{0.400pt}}
\put(506,223.17){\rule{1.700pt}{0.400pt}}
\multiput(506.00,224.17)(4.472,-2.000){2}{\rule{0.850pt}{0.400pt}}
\put(514,221.67){\rule{0.723pt}{0.400pt}}
\multiput(514.00,222.17)(1.500,-1.000){2}{\rule{0.361pt}{0.400pt}}
\put(517,220.17){\rule{1.500pt}{0.400pt}}
\multiput(517.00,221.17)(3.887,-2.000){2}{\rule{0.750pt}{0.400pt}}
\put(466.0,237.0){\rule[-0.200pt]{0.482pt}{0.400pt}}
\put(528,218.17){\rule{1.300pt}{0.400pt}}
\multiput(528.00,219.17)(3.302,-2.000){2}{\rule{0.650pt}{0.400pt}}
\put(524.0,220.0){\rule[-0.200pt]{0.964pt}{0.400pt}}
\put(538,216.17){\rule{1.300pt}{0.400pt}}
\multiput(538.00,217.17)(3.302,-2.000){2}{\rule{0.650pt}{0.400pt}}
\put(534.0,218.0){\rule[-0.200pt]{0.964pt}{0.400pt}}
\put(547,214.67){\rule{1.445pt}{0.400pt}}
\multiput(547.00,215.17)(3.000,-1.000){2}{\rule{0.723pt}{0.400pt}}
\put(553,213.67){\rule{0.723pt}{0.400pt}}
\multiput(553.00,214.17)(1.500,-1.000){2}{\rule{0.361pt}{0.400pt}}
\put(556,212.67){\rule{1.445pt}{0.400pt}}
\multiput(556.00,213.17)(3.000,-1.000){2}{\rule{0.723pt}{0.400pt}}
\put(544.0,216.0){\rule[-0.200pt]{0.723pt}{0.400pt}}
\put(564,211.67){\rule{1.445pt}{0.400pt}}
\multiput(564.00,212.17)(3.000,-1.000){2}{\rule{0.723pt}{0.400pt}}
\put(562.0,213.0){\rule[-0.200pt]{0.482pt}{0.400pt}}
\put(573,210.67){\rule{1.204pt}{0.400pt}}
\multiput(573.00,211.17)(2.500,-1.000){2}{\rule{0.602pt}{0.400pt}}
\put(578,209.67){\rule{0.482pt}{0.400pt}}
\multiput(578.00,210.17)(1.000,-1.000){2}{\rule{0.241pt}{0.400pt}}
\put(580,208.67){\rule{2.409pt}{0.400pt}}
\multiput(580.00,209.17)(5.000,-1.000){2}{\rule{1.204pt}{0.400pt}}
\put(570.0,212.0){\rule[-0.200pt]{0.723pt}{0.400pt}}
\put(595,207.17){\rule{1.900pt}{0.400pt}}
\multiput(595.00,208.17)(5.056,-2.000){2}{\rule{0.950pt}{0.400pt}}
\put(590.0,209.0){\rule[-0.200pt]{1.204pt}{0.400pt}}
\put(609,205.67){\rule{1.927pt}{0.400pt}}
\multiput(609.00,206.17)(4.000,-1.000){2}{\rule{0.964pt}{0.400pt}}
\put(604.0,207.0){\rule[-0.200pt]{1.204pt}{0.400pt}}
\put(621,204.67){\rule{1.927pt}{0.400pt}}
\multiput(621.00,205.17)(4.000,-1.000){2}{\rule{0.964pt}{0.400pt}}
\put(629,203.67){\rule{0.964pt}{0.400pt}}
\multiput(629.00,204.17)(2.000,-1.000){2}{\rule{0.482pt}{0.400pt}}
\put(617.0,206.0){\rule[-0.200pt]{0.964pt}{0.400pt}}
\put(641,202.67){\rule{0.723pt}{0.400pt}}
\multiput(641.00,203.17)(1.500,-1.000){2}{\rule{0.361pt}{0.400pt}}
\put(633.0,204.0){\rule[-0.200pt]{1.927pt}{0.400pt}}
\put(651,201.67){\rule{0.964pt}{0.400pt}}
\multiput(651.00,202.17)(2.000,-1.000){2}{\rule{0.482pt}{0.400pt}}
\put(644.0,203.0){\rule[-0.200pt]{1.686pt}{0.400pt}}
\put(664,200.67){\rule{1.686pt}{0.400pt}}
\multiput(664.00,201.17)(3.500,-1.000){2}{\rule{0.843pt}{0.400pt}}
\put(655.0,202.0){\rule[-0.200pt]{2.168pt}{0.400pt}}
\put(680,199.67){\rule{0.723pt}{0.400pt}}
\multiput(680.00,200.17)(1.500,-1.000){2}{\rule{0.361pt}{0.400pt}}
\put(671.0,201.0){\rule[-0.200pt]{2.168pt}{0.400pt}}
\put(691,198.67){\rule{1.445pt}{0.400pt}}
\multiput(691.00,199.17)(3.000,-1.000){2}{\rule{0.723pt}{0.400pt}}
\put(683.0,200.0){\rule[-0.200pt]{1.927pt}{0.400pt}}
\put(710,197.67){\rule{1.204pt}{0.400pt}}
\multiput(710.00,198.17)(2.500,-1.000){2}{\rule{0.602pt}{0.400pt}}
\put(697.0,199.0){\rule[-0.200pt]{3.132pt}{0.400pt}}
\put(729,196.67){\rule{2.168pt}{0.400pt}}
\multiput(729.00,197.17)(4.500,-1.000){2}{\rule{1.084pt}{0.400pt}}
\put(715.0,198.0){\rule[-0.200pt]{3.373pt}{0.400pt}}
\put(750,195.67){\rule{0.964pt}{0.400pt}}
\multiput(750.00,196.17)(2.000,-1.000){2}{\rule{0.482pt}{0.400pt}}
\put(738.0,197.0){\rule[-0.200pt]{2.891pt}{0.400pt}}
\put(776,194.67){\rule{1.686pt}{0.400pt}}
\multiput(776.00,195.17)(3.500,-1.000){2}{\rule{0.843pt}{0.400pt}}
\put(754.0,196.0){\rule[-0.200pt]{5.300pt}{0.400pt}}
\put(805,193.67){\rule{1.445pt}{0.400pt}}
\multiput(805.00,194.17)(3.000,-1.000){2}{\rule{0.723pt}{0.400pt}}
\put(783.0,195.0){\rule[-0.200pt]{5.300pt}{0.400pt}}
\put(840,192.67){\rule{1.204pt}{0.400pt}}
\multiput(840.00,193.17)(2.500,-1.000){2}{\rule{0.602pt}{0.400pt}}
\put(811.0,194.0){\rule[-0.200pt]{6.986pt}{0.400pt}}
\put(883,191.67){\rule{1.927pt}{0.400pt}}
\multiput(883.00,192.17)(4.000,-1.000){2}{\rule{0.964pt}{0.400pt}}
\put(845.0,193.0){\rule[-0.200pt]{9.154pt}{0.400pt}}
\put(951,190.67){\rule{1.204pt}{0.400pt}}
\multiput(951.00,191.17)(2.500,-1.000){2}{\rule{0.602pt}{0.400pt}}
\put(891.0,192.0){\rule[-0.200pt]{14.454pt}{0.400pt}}
\put(1049,189.67){\rule{0.723pt}{0.400pt}}
\multiput(1049.00,190.17)(1.500,-1.000){2}{\rule{0.361pt}{0.400pt}}
\put(956.0,191.0){\rule[-0.200pt]{22.404pt}{0.400pt}}
\put(1294,188.67){\rule{0.723pt}{0.400pt}}
\multiput(1294.00,189.17)(1.500,-1.000){2}{\rule{0.361pt}{0.400pt}}
\put(1052.0,190.0){\rule[-0.200pt]{58.298pt}{0.400pt}}
\put(1297.0,189.0){\rule[-0.200pt]{33.485pt}{0.400pt}}
\put(176,191){\usebox{\plotpoint}}
\put(176,191){\usebox{\plotpoint}}
\put(176,191){\usebox{\plotpoint}}
\put(176,191){\usebox{\plotpoint}}
\put(176,191){\usebox{\plotpoint}}
\put(176,191){\usebox{\plotpoint}}
\put(176,191){\usebox{\plotpoint}}
\put(176,191){\usebox{\plotpoint}}
\put(176,191){\usebox{\plotpoint}}
\put(176,191){\usebox{\plotpoint}}
\put(176,191){\usebox{\plotpoint}}
\put(176,191){\usebox{\plotpoint}}
\put(176,191){\usebox{\plotpoint}}
\put(176,191){\usebox{\plotpoint}}
\put(176,191){\usebox{\plotpoint}}
\put(176,191){\usebox{\plotpoint}}
\put(176,191){\usebox{\plotpoint}}
\put(176,191){\usebox{\plotpoint}}
\put(176,191){\usebox{\plotpoint}}
\put(176,191){\usebox{\plotpoint}}
\put(176,191){\usebox{\plotpoint}}
\put(176.0,191.0){\usebox{\plotpoint}}
\put(176.0,192.0){\usebox{\plotpoint}}
\put(177.0,192.0){\rule[-0.200pt]{0.400pt}{0.482pt}}
\put(177.0,194.0){\usebox{\plotpoint}}
\put(178.0,194.0){\usebox{\plotpoint}}
\put(178.0,195.0){\rule[-0.200pt]{0.482pt}{0.400pt}}
\put(180.0,195.0){\usebox{\plotpoint}}
\put(180.0,196.0){\usebox{\plotpoint}}
\put(181.0,195.0){\usebox{\plotpoint}}
\put(185,193.67){\rule{0.241pt}{0.400pt}}
\multiput(185.00,194.17)(0.500,-1.000){2}{\rule{0.120pt}{0.400pt}}
\put(181.0,195.0){\rule[-0.200pt]{0.964pt}{0.400pt}}
\put(186,194){\usebox{\plotpoint}}
\put(189,192.67){\rule{0.241pt}{0.400pt}}
\multiput(189.00,193.17)(0.500,-1.000){2}{\rule{0.120pt}{0.400pt}}
\put(186.0,194.0){\rule[-0.200pt]{0.723pt}{0.400pt}}
\put(190,193){\usebox{\plotpoint}}
\put(193,191.67){\rule{0.241pt}{0.400pt}}
\multiput(193.00,192.17)(0.500,-1.000){2}{\rule{0.120pt}{0.400pt}}
\put(190.0,193.0){\rule[-0.200pt]{0.723pt}{0.400pt}}
\put(200,190.67){\rule{0.241pt}{0.400pt}}
\multiput(200.00,191.17)(0.500,-1.000){2}{\rule{0.120pt}{0.400pt}}
\put(194.0,192.0){\rule[-0.200pt]{1.445pt}{0.400pt}}
\put(211,189.67){\rule{0.482pt}{0.400pt}}
\multiput(211.00,190.17)(1.000,-1.000){2}{\rule{0.241pt}{0.400pt}}
\put(201.0,191.0){\rule[-0.200pt]{2.409pt}{0.400pt}}
\put(256,188.67){\rule{0.723pt}{0.400pt}}
\multiput(256.00,189.17)(1.500,-1.000){2}{\rule{0.361pt}{0.400pt}}
\put(213.0,190.0){\rule[-0.200pt]{10.359pt}{0.400pt}}
\put(259.0,189.0){\rule[-0.200pt]{283.539pt}{0.400pt}}
\end{picture}
\caption{This diagram depicts the coefficients $E/A$ 
and ${\cal E}/{\cal A}$
for a typical member of the  
family of the black hole solutions depicted in Fig. 1 ($\phi_h=-1$, 
$r_h=1$);
the coefficient $E/A$ diverges at the horizon as $1/(r-r_h)^2$. 
On the other hand,  
${\cal E}/{\cal A}=e^{\Gamma-\Lambda}E/A$, 
appearing in (40),  
is finite at the horizon. 
The positive-definiteness   
of both coefficients is clear.}
\end{center}
\end{figure}
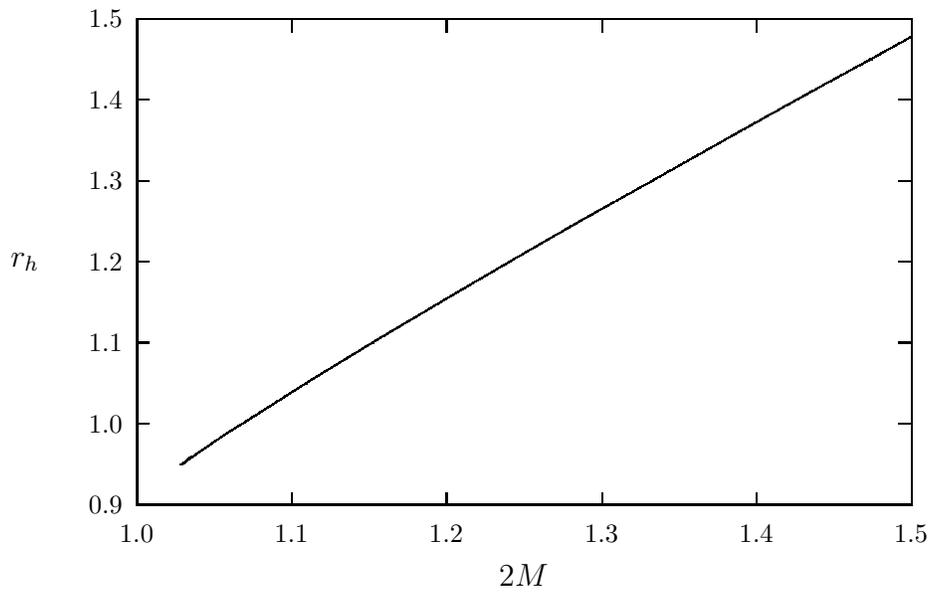
\begin{figure}
\begin{center}
\setlength{\unitlength}{0.240900pt}
\ifx\plotpoint\undefined\newsavebox{\plotpoint}\fi
\begin{picture}(1500,900)(0,0)
\font\gnuplot=cmr10 at 10pt
\gnuplot
\sbox{\plotpoint}{\rule[-0.200pt]{0.400pt}{0.400pt}}%
\put(220.0,113.0){\rule[-0.200pt]{4.818pt}{0.400pt}}
\put(198,113){\makebox(0,0)[r]{0.9}}
\put(1416.0,113.0){\rule[-0.200pt]{4.818pt}{0.400pt}}
\put(220.0,240.0){\rule[-0.200pt]{4.818pt}{0.400pt}}
\put(198,240){\makebox(0,0)[r]{1.0}}
\put(1416.0,240.0){\rule[-0.200pt]{4.818pt}{0.400pt}}
\put(220.0,368.0){\rule[-0.200pt]{4.818pt}{0.400pt}}
\put(198,368){\makebox(0,0)[r]{1.1}}
\put(1416.0,368.0){\rule[-0.200pt]{4.818pt}{0.400pt}}
\put(220.0,495.0){\rule[-0.200pt]{4.818pt}{0.400pt}}
\put(198,495){\makebox(0,0)[r]{1.2}}
\put(1416.0,495.0){\rule[-0.200pt]{4.818pt}{0.400pt}}
\put(220.0,622.0){\rule[-0.200pt]{4.818pt}{0.400pt}}
\put(198,622){\makebox(0,0)[r]{1.3}}
\put(1416.0,622.0){\rule[-0.200pt]{4.818pt}{0.400pt}}
\put(220.0,750.0){\rule[-0.200pt]{4.818pt}{0.400pt}}
\put(198,750){\makebox(0,0)[r]{1.4}}
\put(1416.0,750.0){\rule[-0.200pt]{4.818pt}{0.400pt}}
\put(220.0,877.0){\rule[-0.200pt]{4.818pt}{0.400pt}}
\put(198,877){\makebox(0,0)[r]{1.5}}
\put(1416.0,877.0){\rule[-0.200pt]{4.818pt}{0.400pt}}
\put(220.0,113.0){\rule[-0.200pt]{0.400pt}{4.818pt}}
\put(220,68){\makebox(0,0){1.0}}
\put(220.0,857.0){\rule[-0.200pt]{0.400pt}{4.818pt}}
\put(463.0,113.0){\rule[-0.200pt]{0.400pt}{4.818pt}}
\put(463,68){\makebox(0,0){1.1}}
\put(463.0,857.0){\rule[-0.200pt]{0.400pt}{4.818pt}}
\put(706.0,113.0){\rule[-0.200pt]{0.400pt}{4.818pt}}
\put(706,68){\makebox(0,0){1.2}}
\put(706.0,857.0){\rule[-0.200pt]{0.400pt}{4.818pt}}
\put(950.0,113.0){\rule[-0.200pt]{0.400pt}{4.818pt}}
\put(950,68){\makebox(0,0){1.3}}
\put(950.0,857.0){\rule[-0.200pt]{0.400pt}{4.818pt}}
\put(1193.0,113.0){\rule[-0.200pt]{0.400pt}{4.818pt}}
\put(1193,68){\makebox(0,0){1.4}}
\put(1193.0,857.0){\rule[-0.200pt]{0.400pt}{4.818pt}}
\put(1436.0,113.0){\rule[-0.200pt]{0.400pt}{4.818pt}}
\put(1436,68){\makebox(0,0){1.5}}
\put(1436.0,857.0){\rule[-0.200pt]{0.400pt}{4.818pt}}
\put(220.0,113.0){\rule[-0.200pt]{292.934pt}{0.400pt}}
\put(1436.0,113.0){\rule[-0.200pt]{0.400pt}{184.048pt}}
\put(220.0,877.0){\rule[-0.200pt]{292.934pt}{0.400pt}}
\put(45,495){\makebox(0,0){$r_h$}}
\put(828,3){\makebox(0,0){$2M$}}
\put(220.0,113.0){\rule[-0.200pt]{0.400pt}{184.048pt}}
\multiput(1432.60,846.92)(-0.903,-0.498){67}{\rule{0.820pt}{0.120pt}}
\multiput(1434.30,847.17)(-61.298,-35.000){2}{\rule{0.410pt}{0.400pt}}
\multiput(1369.55,811.92)(-0.914,-0.499){123}{\rule{0.830pt}{0.120pt}}
\multiput(1371.28,812.17)(-113.277,-63.000){2}{\rule{0.415pt}{0.400pt}}
\multiput(1254.60,748.92)(-0.900,-0.499){125}{\rule{0.819pt}{0.120pt}}
\multiput(1256.30,749.17)(-113.301,-64.000){2}{\rule{0.409pt}{0.400pt}}
\multiput(1139.65,684.92)(-0.884,-0.499){125}{\rule{0.806pt}{0.120pt}}
\multiput(1141.33,685.17)(-111.327,-64.000){2}{\rule{0.403pt}{0.400pt}}
\multiput(1026.61,620.92)(-0.898,-0.499){123}{\rule{0.817pt}{0.120pt}}
\multiput(1028.30,621.17)(-111.303,-63.000){2}{\rule{0.409pt}{0.400pt}}
\multiput(913.71,557.92)(-0.869,-0.499){125}{\rule{0.794pt}{0.120pt}}
\multiput(915.35,558.17)(-109.353,-64.000){2}{\rule{0.397pt}{0.400pt}}
\multiput(802.76,493.92)(-0.853,-0.499){125}{\rule{0.781pt}{0.120pt}}
\multiput(804.38,494.17)(-107.378,-64.000){2}{\rule{0.391pt}{0.400pt}}
\multiput(693.79,429.92)(-0.843,-0.499){123}{\rule{0.773pt}{0.120pt}}
\multiput(695.40,430.17)(-104.396,-63.000){2}{\rule{0.387pt}{0.400pt}}
\multiput(587.86,366.92)(-0.821,-0.499){125}{\rule{0.756pt}{0.120pt}}
\multiput(589.43,367.17)(-103.430,-64.000){2}{\rule{0.378pt}{0.400pt}}
\multiput(482.99,302.92)(-0.782,-0.499){125}{\rule{0.725pt}{0.120pt}}
\multiput(484.50,303.17)(-98.495,-64.000){2}{\rule{0.363pt}{0.400pt}}
\multiput(382.82,238.92)(-0.841,-0.492){21}{\rule{0.767pt}{0.119pt}}
\multiput(384.41,239.17)(-18.409,-12.000){2}{\rule{0.383pt}{0.400pt}}
\multiput(363.16,226.92)(-0.734,-0.493){23}{\rule{0.685pt}{0.119pt}}
\multiput(364.58,227.17)(-17.579,-13.000){2}{\rule{0.342pt}{0.400pt}}
\multiput(344.16,213.92)(-0.734,-0.493){23}{\rule{0.685pt}{0.119pt}}
\multiput(345.58,214.17)(-17.579,-13.000){2}{\rule{0.342pt}{0.400pt}}
\multiput(325.16,200.92)(-0.734,-0.493){23}{\rule{0.685pt}{0.119pt}}
\multiput(326.58,201.17)(-17.579,-13.000){2}{\rule{0.342pt}{0.400pt}}
\put(307,187.67){\rule{0.482pt}{0.400pt}}
\multiput(308.00,188.17)(-1.000,-1.000){2}{\rule{0.241pt}{0.400pt}}
\put(305,186.67){\rule{0.482pt}{0.400pt}}
\multiput(306.00,187.17)(-1.000,-1.000){2}{\rule{0.241pt}{0.400pt}}
\put(303,185.67){\rule{0.482pt}{0.400pt}}
\multiput(304.00,186.17)(-1.000,-1.000){2}{\rule{0.241pt}{0.400pt}}
\put(301,184.17){\rule{0.482pt}{0.400pt}}
\multiput(302.00,185.17)(-1.000,-2.000){2}{\rule{0.241pt}{0.400pt}}
\put(299,182.67){\rule{0.482pt}{0.400pt}}
\multiput(300.00,183.17)(-1.000,-1.000){2}{\rule{0.241pt}{0.400pt}}
\put(298,181.67){\rule{0.241pt}{0.400pt}}
\multiput(298.50,182.17)(-0.500,-1.000){2}{\rule{0.120pt}{0.400pt}}
\put(296,180.17){\rule{0.482pt}{0.400pt}}
\multiput(297.00,181.17)(-1.000,-2.000){2}{\rule{0.241pt}{0.400pt}}
\put(294,178.67){\rule{0.482pt}{0.400pt}}
\multiput(295.00,179.17)(-1.000,-1.000){2}{\rule{0.241pt}{0.400pt}}
\put(292,177.67){\rule{0.482pt}{0.400pt}}
\multiput(293.00,178.17)(-1.000,-1.000){2}{\rule{0.241pt}{0.400pt}}
\put(292,178){\usebox{\plotpoint}}
\put(291,176.67){\rule{0.241pt}{0.400pt}}
\multiput(291.50,177.17)(-0.500,-1.000){2}{\rule{0.120pt}{0.400pt}}
\put(291,177){\usebox{\plotpoint}}
\put(290.0,177.0){\usebox{\plotpoint}}
\put(290.0,176.0){\usebox{\plotpoint}}
\put(289.0,176.0){\usebox{\plotpoint}}
\end{picture}
\caption{The graph depicts $r_h$ versus the ADM Mass $2M$
of the black hole, for a fixed value of
$\phi _h=-1$.
The emergence of the asymptotic critical point,
$r_h^4 \simeq 6\alpha^{'2}e^{2\phi _h}/g^4$,
below which there are no solutions,
is apparent. At this point the mass
becomes minimal.}
\end{center}
\end{figure}
\begin{figure}
\begin{center}
\setlength{\unitlength}{0.240900pt}
\ifx\plotpoint\undefined\newsavebox{\plotpoint}\fi
\sbox{\plotpoint}{\rule[-0.200pt]{0.400pt}{0.400pt}}%
\begin{picture}(1500,900)(0,0)
\font\gnuplot=cmr10 at 10pt
\gnuplot
\sbox{\plotpoint}{\rule[-0.200pt]{0.400pt}{0.400pt}}%
\put(220.0,113.0){\rule[-0.200pt]{4.818pt}{0.400pt}}
\put(198,113){\makebox(0,0)[r]{0.948}}
\put(1416.0,113.0){\rule[-0.200pt]{4.818pt}{0.400pt}}
\put(220.0,304.0){\rule[-0.200pt]{4.818pt}{0.400pt}}
\put(198,304){\makebox(0,0)[r]{0.949}}
\put(1416.0,304.0){\rule[-0.200pt]{4.818pt}{0.400pt}}
\put(220.0,495.0){\rule[-0.200pt]{4.818pt}{0.400pt}}
\put(198,495){\makebox(0,0)[r]{0.950}}
\put(1416.0,495.0){\rule[-0.200pt]{4.818pt}{0.400pt}}
\put(220.0,686.0){\rule[-0.200pt]{4.818pt}{0.400pt}}
\put(198,686){\makebox(0,0)[r]{0.951}}
\put(1416.0,686.0){\rule[-0.200pt]{4.818pt}{0.400pt}}
\put(220.0,877.0){\rule[-0.200pt]{4.818pt}{0.400pt}}
\put(198,877){\makebox(0,0)[r]{0.952}}
\put(1416.0,877.0){\rule[-0.200pt]{4.818pt}{0.400pt}}
\put(220.0,113.0){\rule[-0.200pt]{0.400pt}{4.818pt}}
\put(220,68){\makebox(0,0){1.00}}
\put(220.0,857.0){\rule[-0.200pt]{0.400pt}{4.818pt}}
\put(382.0,113.0){\rule[-0.200pt]{0.400pt}{4.818pt}}
\put(382,68){\makebox(0,0){1.02}}
\put(382.0,857.0){\rule[-0.200pt]{0.400pt}{4.818pt}}
\put(544.0,113.0){\rule[-0.200pt]{0.400pt}{4.818pt}}
\put(544,68){\makebox(0,0){1.04}}
\put(544.0,857.0){\rule[-0.200pt]{0.400pt}{4.818pt}}
\put(706.0,113.0){\rule[-0.200pt]{0.400pt}{4.818pt}}
\put(706,68){\makebox(0,0){1.06}}
\put(706.0,857.0){\rule[-0.200pt]{0.400pt}{4.818pt}}
\put(869.0,113.0){\rule[-0.200pt]{0.400pt}{4.818pt}}
\put(869,68){\makebox(0,0){1.08}}
\put(869.0,857.0){\rule[-0.200pt]{0.400pt}{4.818pt}}
\put(1031.0,113.0){\rule[-0.200pt]{0.400pt}{4.818pt}}
\put(1031,68){\makebox(0,0){1.1}}
\put(1031.0,857.0){\rule[-0.200pt]{0.400pt}{4.818pt}}
\put(1193.0,113.0){\rule[-0.200pt]{0.400pt}{4.818pt}}
\put(1193,68){\makebox(0,0){1.12}}
\put(1193.0,857.0){\rule[-0.200pt]{0.400pt}{4.818pt}}
\put(1355.0,113.0){\rule[-0.200pt]{0.400pt}{4.818pt}}
\put(1355,68){\makebox(0,0){1.14}}
\put(1355.0,857.0){\rule[-0.200pt]{0.400pt}{4.818pt}}
\put(220.0,113.0){\rule[-0.200pt]{292.934pt}{0.400pt}}
\put(1436.0,113.0){\rule[-0.200pt]{0.400pt}{184.048pt}}
\put(220.0,877.0){\rule[-0.200pt]{292.934pt}{0.400pt}}
\put(45,495){\makebox(0,0){$r_h$}}
\put(828,3){\makebox(0,0){$2M$}}
\put(220.0,113.0){\rule[-0.200pt]{0.400pt}{184.048pt}}
\put(466,877){\usebox{\plotpoint}}
\multiput(464.93,823.73)(-0.482,-17.216){9}{\rule{0.116pt}{12.833pt}}
\multiput(465.17,850.36)(-6.000,-164.364){2}{\rule{0.400pt}{6.417pt}}
\put(458.67,648){\rule{0.400pt}{9.154pt}}
\multiput(459.17,667.00)(-1.000,-19.000){2}{\rule{0.400pt}{4.577pt}}
\put(457.17,610){\rule{0.400pt}{7.700pt}}
\multiput(458.17,632.02)(-2.000,-22.018){2}{\rule{0.400pt}{3.850pt}}
\put(455.67,571){\rule{0.400pt}{9.395pt}}
\multiput(456.17,590.50)(-1.000,-19.500){2}{\rule{0.400pt}{4.698pt}}
\put(454.67,533){\rule{0.400pt}{9.154pt}}
\multiput(455.17,552.00)(-1.000,-19.000){2}{\rule{0.400pt}{4.577pt}}
\put(453.67,495){\rule{0.400pt}{9.154pt}}
\multiput(454.17,514.00)(-1.000,-19.000){2}{\rule{0.400pt}{4.577pt}}
\put(452.67,476){\rule{0.400pt}{4.577pt}}
\multiput(453.17,485.50)(-1.000,-9.500){2}{\rule{0.400pt}{2.289pt}}
\put(451.67,457){\rule{0.400pt}{4.577pt}}
\multiput(452.17,466.50)(-1.000,-9.500){2}{\rule{0.400pt}{2.289pt}}
\put(450.67,419){\rule{0.400pt}{4.577pt}}
\multiput(451.17,428.50)(-1.000,-9.500){2}{\rule{0.400pt}{2.289pt}}
\put(452.0,438.0){\rule[-0.200pt]{0.400pt}{4.577pt}}
\put(449.67,380){\rule{0.400pt}{4.818pt}}
\multiput(450.17,390.00)(-1.000,-10.000){2}{\rule{0.400pt}{2.409pt}}
\put(448.67,361){\rule{0.400pt}{4.577pt}}
\multiput(449.17,370.50)(-1.000,-9.500){2}{\rule{0.400pt}{2.289pt}}
\put(451.0,400.0){\rule[-0.200pt]{0.400pt}{4.577pt}}
\put(449.0,356.0){\rule[-0.200pt]{0.400pt}{1.204pt}}
\end{picture}
\caption{
The magnification around the critical point, depicted in the above
figure, shows clearly the abrupt (almost vertical) slope
with which this point is approached. No turning point is found.}
\end{center}
\end{figure}
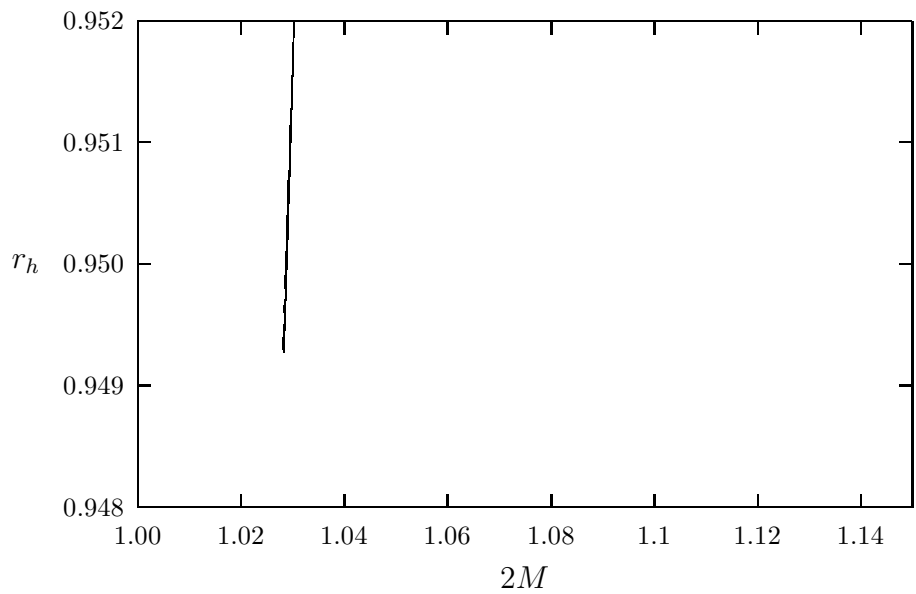
\end{document}